\documentclass[aps,prd,onecolumn,eqsecnum,amsmath,nofootinbib,preprintnumbers]{revtex4}%

\usepackage{color,graphicx,float,subfigure}
\usepackage{amsfonts,amssymb,theorem,mathrsfs,times}
\usepackage{bm}
\usepackage{mathtools}
\usepackage{amsfonts,amssymb,theorem,mathrsfs}
\usepackage{dsfont}
\usepackage{setspace}

\usepackage[colorlinks,linkcolor=blue,anchorcolor=blue,citecolor=blue]{hyperref}   
\usepackage{ulem}   

{\theorembodyfont{\upshape}
}
{\theorembodyfont{\upshape}
}
{\theorembodyfont{\upshape}
}
{\theorembodyfont{\upshape}
}
{\theorembodyfont{\upshape}
}
{\theorembodyfont{\upshape}
}
\addtocounter{MaxMatrixCols}{10}
\newcommand{\dalm}{\kern1pt\vbox{\hrule height 0.9pt\hbox{\vrule width
0.9pt\hskip 2.5pt\vbox{\vskip 5.5pt}\hskip 3pt\vrule width
0.3pt}\hrule height 0.3pt}\kern1pt}

\begin{document}
\preprint{\hfill {\small {ICTS-USTC/PCFT-21-47}}}
\title{A Note on the Strong Hyperbolicity of $f(R)$ Gravity with Dynamical Shifts}

%

\author{ Li-Ming Cao$^{a\, ,b}$\footnote{e-mail
address: caolm@ustc.edu.cn}}

\author{ Liang-Bi Wu$^b$\footnote{e-mail
address: liangbi@mail.ustc.edu.cn}}

\affiliation{$^a$Peng Huanwu Center for Fundamental Theory, Hefei, Anhui 230026, China}

\affiliation{${}^b$
Interdisciplinary Center for Theoretical Study and Department of Modern Physics,\\
University of Science and Technology of China, Hefei, Anhui 230026,
China}

\date{\today}

\begin{abstract}

The well-posedness of the gravitational equations of $f(R)$ gravity are studied in this paper. Three formulations of the $f(R)$ gravity with  dynamical shifts (which are all based on the Arnowitt-Deser-Misner (ADM) formalism of the equations) are investigated. These three formulations are all proved to be strongly hyperbolic by pseudodifferential reduction. The first one is the  Baumagarte-Shapiro-Shibata-Nakamura (BSSN) formulation with the so-called ``hyperbolic $K$-driver" condition and the ``hyperbolic Gamma driver" condition. The second one is the ADM formulation with modified harmonic gauge conditions. We find that the equations are not strong hyperbolic in traditional Z4 formulation for $f(R)$ gravity. So, in the third formulation, we improve the Z4 formulation, and show these equations are strong hyperbolic with modified harmonic gauge conditions.

\end{abstract}


\maketitle

\section{Introduction}\label{section1}

In the past two decades, physicists have been trying to find an interpretation about the early and late time accelerating expansion of the universe. As an alternative solution, modified gravity has been arousing people's curiosities. The so-called $f(R)$ gravity, whose Lagrangian is an analytic function of the spacetime's Ricci scalar, is one of the simplest and the most direct modifications of general relativity. Starobinsky used the model with $f(R)=R+\alpha R^2$ to give an explanation of the early accelerating expansion of the universe~\cite{Starobinsky:1980te}, without introducing extra inflation fields. Gradually, the investigations of $f(R)$ gravity is expanded to many aspects, ranging from solar system to cosmology~\cite{Li:2007xn,Lecian:2008vc}. To get a systematic understanding of $f(R)$ gravity, one can read two nice reviews~\cite{Sotiriou:2008rp,DeFelice:2010aj}.

However, there is still much to be studied about $f(R)$ gravity such as its well-posed initial value problem (IVP). A well-posed initial value problem in some sense has the following three reasonable properties associated with the equations of motion. Given suitable initial data and boundary data, (i) a solution must be existent, (ii) the solution must be unique, and (iii) the solution must depend continuously on the initial data. The well-posed initial value problem has been successfully demonstrated in general relativity, which enables us to make predictions under strong field or dynamical field condition, with the powerful tool of numerical relativity. What's more, the well-posedness also demonstrates the local determinism  of classical theories. Therefore, naturally, we also expect $f(R)$ gravity to have a well-posed initial value problem. In addition to the $f(R)$ gravity, the well-posed formulations in other modified gravities have been put forward. Scalar-tensor theory has been discussed in Ref.\cite{Salgado:2008xh}. For the Einstein-\ae ther theory, the well-posed formulation is given in Ref.\cite{Sarbach:2019yso}, where the authors use the Ricci rotation coefficients (Such a formulation was obtained in general relativity \cite{Buchman:2003sq}.). The well-posed formulation of cubic Horndeski theories is proposed in Ref.\cite{Kovacs:2019jqj}. 

When we discuss a well-posed initial value problem, the concept of hyperbolicity is resorted. Hyperbolicity refers to algebraic conditions on the principal part of the equations. It implies well-posedness for the Cauchy problem, which reveals the existence of a unique continuous map between initial data and solutions. Especially, among a series of the definitions of hyperbolicity, strong hyperbolicity is consider to be a spot-on definition for a well-posed initial value problem. The proof of this equivalence is based on pseudodiffenrential analysis~\cite{Taylor1,Taylor2}. 

Roughly speaking, there are two mainstream methods to investigate the hyperbolicity of a gravity theory. One is based on the the linearized equations of motion. The well-posed formulation in Horndeski and Lovelock gravity at the perturbative level has been studied in~\cite{Papallo:2017qvl,Papallo:2017ddx,Kovacs:2020ywu,Cao:2021sty,Cao:2021nng}. Certainly, it is important for us to find a well-posed, nonperturbative formulation in general relativity or modified gravities. So, the other is nonperturbative. It is based on the Arnowitt-Deser-Misner (ADM) decomposition of the evolution equations~\cite{Arnowitt:1962hi} with some suitable gauge conditions or the Bondi-Sachs formalism~\cite{Bondi:1962px,Winicour:2008vpn,Giannakopoulos:2020dih}. The key technology of strong hyperbolicity is to check whether the eigenvalues of the principal part are all real or not and check whether the eigenvectors of the principal part of the equations span the whole eigenspace or not~\cite{Nagy:2004td,Sarbach:2012pr,1989Initial}. 

It has been proved that the ADM evolution equations are of weak hyperbolicity in general relativity~\cite{Kidder:2001tz}. This is the reason why one can find some instabilities in ADM formulations~\cite{Calabrese:2002ej,Calabrese:2002ei,Baumgarte:1998te}. In the Ref.\cite{Nagy:2004td}, the authors came up with densitied ADM equations where the lapse function is densitized. However, this formulation is still not strongly hyperbolic but only weakly hyperbolic. Baumgarate-Shapiro-Shibata-Nakamura (BSSN) type systems are considered to solve the problem of strong hyperbolic destruction in the ADM formulation. It is based on the ADM decomposition of the field equations. A new variable $\tilde{\Gamma}^i$ is introduced in the BSSN systems. It has been showed in many papers that the BSSN formulation leads to strong hyperbolicity of the evolution equations in general relativity~\cite{Sarbach:2002bt,Beyer:2004sv,Reula:2004xd}. 

It's worth pointing out that all notions of hyperbolicity mentioned above require that the evolution equations are first order systems. Here, either the ADM evolution equations or the BSSN evolution equations are first order in time, but mixed first/second order in space. Hence, the strategy to analyze the hyperbolicity of these second-order systems is to transform them into equivalent first-order systems. Then one looks at algebraic properties of the principal part for these first-order systems. There are several ways of obtaining a first-order system from these second-order ones. One of them, used in~\cite{Sarbach:2002bt,Beyer:2004sv,Frittelli:1999sj}, is to add as variables all first-order derivatives and look at the resulting larger system. Another is to add as new variables the square root of the Laplacian of some of the original variables and so get a first-order pseudo-differential system~\cite{Reula:2004xd}. No extra equations will be introduced under the pseudo-differential reductions which were first used in general relativity in~\cite{Kriess}.

In the early days, the Cauchy problem of $f(R)$ gravity was studied through the equivalency of $f(R)$ gravity and scalar-tensor gravity~\cite{Lanahan-Tremblay:2007sxd}. In 2016, Mongwane, bypassing this equivalency, took full advantage of the ADM decomposition proposed in Ref.\cite{Tsokaros:2013fma} to directly study the hyperbolicty of the $f(R)$ gravity. He investigated the hyperbolicity of $f(R)$ gravity but only for a given shift function in both the ADM formulation and the BSSN formulation. Therefore, by adding all first-order derivatives as variables, he proved the ADM version of $f(R)$ gravity is just weakly hyperbolic while the BSSN version of $f(R)$ gravity is strongly hyperbolic~\cite{Mongwane:2016qtz}. 

On the one hand, it is known that there are many other formulations with different gauge conditions instead of the BSSN formulation with Bona-Masso slicing condition~\cite{Bona:1994dr} which is used in the Ref.\cite{Mongwane:2016qtz} for the $f(R)$ gravity. For example, ADM formulation with harmonic gauge conditions and the Z4 formulation are common formulations. Then, it is natural for us to study whether or not these formulations keep the strong hyperbolicity in $f(R)$ gravity. On the other hand, the advantage of the pseudodiffenrential reduction is that the principal part of the system is algebraically much simpler to dispose of (the matrix of principal part is much smaller), especially for systems who have second order derivatives in space. Based on these two reasons, the main purpose of this paper is to make an investigation of the hyperbolicity of $f(R)$ gravity in three different formulations by using the pseudodiffenrential reduction. To be specific, the first one considered is the BSSN formulation with dynamical shifts and lapses, we find it will be strongly hyperbolic under the so-called ``hyperbolic $K$-driver" condition and the ``hyperbolic Gamma driver" condition. The second one is the ADM formulation with modified harmonic gauge condition which is different from the one in general relativity~\cite{Sarbach:2012pr}. For the last formulation, a 4-vector field $\mathcal{Z}^a$ is added into the original equations of motion. Then we get the so-called Z4 formulation. However, the approach  to add $\mathcal{Z}^a$ is distinct with the traditional one in general relativity. Otherwise, one cannot acquire a strongly hyperbolic Z4 formulation.

This paper is organized as follows. In Sec.\ref{section2}, we present the standard Arnowitt-Deser-Misner (ADM) formulation of $f(R)$ gravity according to the Ref.\cite{Tsokaros:2013fma}. The hyperbolicity of the Baumgarte-Shapiro-Shibata-Nakamura (BSSN) formulation with dynamical lapse and dynamical shift is studied in Sec.\ref{section3}. In Sec.\ref{section4}, a modified harmonic formulation is analyzed. Last but not least, we perform a modified Z4 formulation for the $f(R)$ gravity in Sec.\ref{section5}. Sec.\ref{section6} is the conclusion and the discussion.

\section{The ADM decomposition of equations of motion in $f(R)$ gravity}\label{section2}
We start by a brief review of the ADM decomposition of equations of motion in $f(R)$ gravity (the details can be found in Ref.\cite{Tsokaros:2013fma}). 
There are three versions of $f(R)$ gravity, metric formalism, Palatini formalism, and metric-affine gravity, respectively~\cite{Sotiriou:2006hs}. 
In this paper, we consider the metric formalism whose action can be expressed as
\begin{eqnarray}\label{action}
	S=\frac{1}{2\kappa^2}\int\mathrm{d}^4x\Big[\sqrt{-g}f(R)+2\kappa^2\mathcal{L}_m\Big]\, ,
\end{eqnarray}
where $\kappa^2=8\pi G_N$, and $G_N$ is the gravitational constant. The symbol $g$ is the determinant of the spacetime metric $g_{ab}$, and $\mathcal{L}_m$ is the Lagrangian density for usual matter fields.
Varying the action (\ref{action}) with respect to the metric $g_{ab}$ yields equations of motion
\begin{eqnarray}\label{EOM}
	f^{\prime}R_{ab}-\frac{1}{2}fg_{ab}-\nabla_a\nabla_bf^{\prime}+g_{ab}\square f^{\prime}=\kappa^2T_{ab}\, ,
\end{eqnarray}
where $f^{\prime}=\partial f(R)/\partial R$, $f=f(R)$, and $T_{ab}$ is the energy-momentum tensor  from the Lagrangian density $\mathcal{L}_m$.
For convenience on the notion, usually, we introduce a symmetric tensor $\Sigma_{ab}$ which is defined as follows
\begin{eqnarray}
	\Sigma_{ab}\equiv f^{\prime}R_{ab}-\frac{1}{2}fg_{ab}-f^{\prime\prime}\nabla_a\nabla_bR-f^{\prime\prime\prime}\nabla_aR\nabla_bR+g_{ab}(f^{\prime\prime\prime}\nabla^cR\nabla_cR+f^{\prime\prime}\square R)\, .
\end{eqnarray}
Hence, Eq.(\ref{EOM}) can be written sententiously as
\begin{eqnarray}
	\Sigma_{ab}=\kappa^2T_{ab}\, .
\end{eqnarray}
We think of spacetime $(M,g_{ab})$ to be foliated by spacelike surfaces $(\Sigma_t\, ,\gamma_{ab})$. 
Let $n^a$ be the future directed unit normal to $\Sigma_t$ and the induced spatial metric on those hypersurfaces is
\begin{eqnarray}
	\gamma_{ab}=g_{ab}+n_an_b\, .
\end{eqnarray}
The mixed tensor $\gamma_a{}^b$ is called the projection operator since when contracted with any four-dimensional vector it produces its spatial projection on $\Sigma_t$. 
Under the standard ADM decomposition, the metric $g_{ab}$ is written as
\begin{eqnarray}\label{metric}
	\mathrm{d}s^2=-(\alpha^2-\beta_i\beta^i)\mathrm{d}t^2+2\beta_i\mathrm{d}t
	\mathrm{d}x^i+\gamma_{ij}\mathrm{d}x^i\mathrm{d}x^j\, ,
\end{eqnarray}
where $\alpha$ is the lapse function and $\beta^i$ is the shift vector with $\beta_i=\gamma_{ij}\beta^j$.  The induced covariant derivative on $(\Sigma_t\, ,\gamma_{ab})$ is denoted by $D_a$,  
which  is compatible with the induced metric $\gamma_{ab}$ as usual. The extrinsic curvature $K_{ab}$ can be defined in terms of projections of covariant derivative of $n_a$, i.e.,
\begin{eqnarray}\label{Kab}
	K_{ab}&=&-\gamma_a{}^c\gamma_b{}^d\nabla_cn_d=-\nabla_an_b-n_aa_b=-\frac12\mathcal{L}_n\gamma_{ab}\, ,
\end{eqnarray}
where $a_b=n^c\nabla_cn_b$ is the acceleration of the normal $n_a$, and it is related to the lapse function $\alpha$ via $a_b=D_b\ln\alpha$.
By these definitions, the ADM decomposition of the full system is expressed as follows~\cite{Tsokaros:2013fma,Mongwane:2016qtz}
\begin{eqnarray}
	\partial_0R&=&\alpha\psi\, ,\label{ADM_R}\\
	\partial_0\psi&=&\frac{\alpha}{3f^{\prime\prime}}\Big[-2f+Rf^{\prime}+3\Big(D_iD^iR+K\psi+a^iD_iR\Big)f^{\prime\prime}\nonumber\\
	&&+3\Big(D^iRD_iR-\psi^2\Big)f^{\prime\prime\prime}-\kappa^2(S-\rho)\Big]\, ,\label{ADM_psi}\\
	\partial_0\gamma_{ij}&=&-2\alpha K_{ij}+\gamma_{ik}\partial_j\beta^k+\gamma_{jk}\partial_i\beta^k\, ,\label{ADM_gammaij}\\
	\partial_0K_{ij}&=&\alpha\Big(\mathcal{R}_{ij}-2K_i{}^kK_{jk}+KK_{ij}\Big)+\frac{\alpha}{f^{\prime}}\left\lbrace\frac16f\gamma_{ij}-\frac13\gamma_{ij}Rf^{\prime}\right.\nonumber\\
	&&\left.-\Big(D_iD_jR+\psi K_{ij}\Big)f^{\prime\prime}-f^{\prime\prime\prime}D_iRD_jR-\kappa^2\Big[S_{ij}-\frac13\gamma_{ij}(S-\rho)\Big]\right\rbrace\nonumber\\
	&&-D_iD_j\alpha+K_{kj}\partial_i\beta^k+K_{ik}\partial_j\beta^k\, \label{ADM_Kij}.
\end{eqnarray}
Here, the operator $\partial_0$ is defined as $\partial_0\equiv\partial_t-\beta^i\partial_i$ with $\beta^i$ denoting shift. 
This operator is always used in the following discussion. In the above equations, $\mathcal{R}_{ij}$ called the spatial Ricci tensor is obtained by the following expression
\begin{eqnarray}
	\mathcal{R}_{ij}=\frac12\gamma^{kl}\Big(\partial_i\partial_l\gamma_{kj}+\partial_k\partial_j\gamma_{il}-\partial_i\partial_j\gamma_{kl}-\partial_k\partial_l\gamma_{ij}\Big)+\gamma^{kl}\Big(\Gamma^m{}_{il}\Gamma_{mkj}-\Gamma^m{}_{ij}\Gamma_{mkl}\Big)\, ,
\end{eqnarray}
where $\Gamma$ is computed from $\gamma_{ij}$, $\psi$ is defined as the Lie derivative of $R$ along $n$, and $K$ is the trace of $K_{ij}$. The quantities $\rho$, $S_{ij}$, $S$ come from the energy-momentum tensor $T_{ab}$, i.e.,
\begin{eqnarray}
	\rho=n^an^bT_{ab}\, ,\quad S_{cd}=\gamma^a{}_c\gamma^b{}_dT_{ab}\, ,\quad S=\gamma^{ab}T_{ab}\, .
\end{eqnarray}
It should be noted here, for $f(R)$ gravity, that the evolution variables are $\left\lbrace R, \psi, \gamma_{ij}, K_{ij}\right\rbrace$, 
which is very different  from the one in general relativity in which the variables of evolution are merely $\left\lbrace\gamma_{ij}, K_{ij}\right\rbrace$. 
If we do not think of $R$ as an independent dynamical variable, we have to consider evolution equations whose highest derivative is quartic but not quadratic.

\section{BSSN formulation with dynamical lapse and dynamical shift}\label{section3}

In general relativity, we know the ADM formulation are not strongly hyperbolic~\cite{Nagy:2004td}. But for the BSSN formulation, the strong hyperbolicity will be hold~\cite{Beyer:2004sv,Nagy:2004td}. 
Moreover, in the case of $f(R)$, the ADM equations are still not strongly hyperbolic, while the relevant BSSN formulation will keep the strongly hyperbolicity when one chooses suitable parameters~\cite{Mongwane:2016qtz}. 
The BSSN formulation is based on the ADM formulation. In this subsection, we will show how the BSSN formulation keep the strong hyperbolicity in $f(R)$ gravity. 
However, unlike the method used in the Ref.\cite{Mongwane:2016qtz} with a fixed shift vector, we will use pseudodifferential reductions to complete the analysis under the dynamical lapse and the dynamical shift.

First, we write down the BSSN formulation explicitly for $f(R)$ gravity which was proposed in the Ref.\cite{Mongwane:2016qtz}. 
In BSSN formulation, the three metric $\gamma_{ij}$ and the extrinsic curvature $K_{ij}$ are decomposed according to
\begin{eqnarray}
	\gamma_{ij}&=&e^{4\phi}\tilde{\gamma}_{ij}\, ,\label{decomposition_gamma}\\
	K_{ij}&=&e^{4\phi}\Big(\tilde{A}_{ij}+\frac13\tilde{\gamma}_{ij}K\Big)\, ,\label{decomposition_Kij}
\end{eqnarray}
where $\tilde{\gamma}_{ij}$ has unit determinant. A new variable defined as 
\begin{eqnarray}
	\tilde{\Gamma}^i\equiv\tilde{\gamma}^{jk}\tilde{\Gamma}^i{}_{jk}
\end{eqnarray}
is added in the BSSN formulation. In summary, the variables of evolution for $f(R)$ gravity are $$\{ R, \psi, \phi, K, \tilde{\Gamma}^i, \tilde{\gamma}_{ij}, \tilde{A}_{ij}\}\, .$$ 
The evolution equations of the BSSN formulation for $f(R)$ gravity have following forms
\begin{eqnarray}
	\partial_0R&=&\alpha\psi\, ,\\
	\partial_0\psi&=&\frac{\alpha}{3f^{\prime\prime}}\Big[-2f+Rf^{\prime}+3\Big(D_iD^iR+K\psi+a^iD_iR\Big)f^{\prime\prime}\nonumber\\
	&&+3\Big(D^iRD_iR-\psi^2\Big)f^{\prime\prime\prime}-\kappa^2(S-\rho)\Big]\, ,\\
	\partial_0\phi&=&-\frac{1}{6}\alpha K+\frac16\partial_k\beta^k\, ,\\
	\partial_0\tilde{\gamma}_{ij}&=&-2\alpha\tilde{A}_{ij}+\tilde{\gamma}_{ik}\partial_j\beta^k+\tilde{\gamma}_{jk}\partial_i\beta^k-\frac23\tilde{\gamma}_{ij}\partial_k\beta^k\, ,\\
	\partial_0K&=&\frac{\alpha}{f^{\prime}}\Big[-\frac12f+f^{\prime\prime}\Big(D^iD_iR+K\psi\Big)+f^{\prime\prime\prime}D^iRD_iR+\kappa^2\rho\Big]\nonumber\\
	&&+\alpha\Big(\tilde{A}_{ij}\tilde{A}^{ij}+\frac13K^2\Big)-\gamma^{ij}D_iD_j\alpha\, ,\\
	\partial_0\tilde{A}_{ij}&=&\alpha\Big(K\tilde{A}_{ij}-\tilde{A}_{ik}\tilde{A}^k{}_j\Big)+e^{-4\phi}\Big(\alpha\mathcal{R}_{ij}-D_iD_j\alpha\Big)^{\text{TF}}\nonumber\\
	&&-\frac{\alpha e^{-4\phi}}{f^{\prime}}\left\lbrace\Big[D_iD_jR+\psi e^{4\phi}\Big(\tilde{A}_{ij}+\frac13K\tilde{\gamma}_{ij}\Big)\Big]f^{\prime\prime}+f^{\prime\prime\prime}D_iRD_jR+\kappa^2S_{ij}\right\rbrace^{\text{TF}}\nonumber\\
	&&+\tilde{A}_{ik}\partial_j\beta^k+\tilde{A}_{jk}\partial_i\beta^k-\frac23\tilde{A}_{ij}\partial_k\beta^k\, ,\\
	\partial_0\tilde{\Gamma}^i&=&2\alpha\tilde{\Gamma}^i{}_{jk}\tilde{A}^{jk}-\frac43\alpha m\tilde{\gamma}^{ij}D_jK+12\alpha m\tilde{A}^{ij}D_j\phi+2\alpha(m-1)\tilde{D}_j\tilde{A}^{ij}-2\tilde{A}^{ij}D_j\alpha\nonumber\\
	&&+\tilde{\gamma}^{jk}\partial_j\partial_k\beta^i+\frac13\tilde{\gamma}^{ij}\partial_j\partial_k\beta^k-\tilde{\Gamma}^j\partial_j\beta^i+\frac23\tilde{\Gamma}^i\partial_j\beta^j-2\alpha m\kappa^2e^{4\phi}\frac{S^i}{f^{\prime}}\nonumber\\
	&&+2\alpha m\frac{f^{\prime\prime\prime}}{f^{\prime}}\tilde{\gamma}^{ij}\psi D_jR+2\alpha m\frac{f^{\prime\prime}}{f^{\prime}}\Big[\Big(\tilde{A}^{ij}+\frac{1}{3}\tilde{\gamma}^{ij}K\Big)D_jR+\tilde{\gamma}^{ij}D_j\psi\Big]\, ,\label{BSSN_Gamma}
\end{eqnarray}
where $\tilde{D}_i$ is compatible with the conformal metric $\tilde{\gamma}_{ij}$, and $S_c=-\gamma^a{}_cn^bT_{ab}$ is  the momentum density of the matter fields. It should be noted that the indices of quantities which have a ``tilde" are lowered and raised by the conformal metric $\tilde{\gamma}_{ij}$. The expression $[\dots]^{\text{TF}}$ denotes the traceless part of terms inside the square brackets with respect to the metric $\tilde{\gamma}_{ij}$. The parameter $m$ introduced in~\cite{Beyer:2004sv, Alcubierre:1999rt}, manifests how the momentum constraint is added to the evolution equations for the variable $\tilde{\Gamma}^i$, where the momentum constraint is expressed as~\cite{Mongwane:2016qtz}
\begin{eqnarray}
	\kappa^2e^{4\phi}S^i-\Big(\tilde{D}_j\tilde{A}^{ij}+6\tilde{A}^{ij}\tilde{D}_j\phi-\frac23\tilde{\gamma}^{ij}D_jK\Big)f^{\prime}-f^{\prime\prime}\Big[\Big(\tilde{A}^{ij}+\frac13\tilde{\gamma}^{ij}K\Big)D_jR+\tilde{\gamma}^{ij}D_j\psi\Big]-\tilde{\gamma}^{ij}f^{\prime\prime\prime}\psi D_jR=0\, .
\end{eqnarray}
Here, we consider the gauge condition given in the Ref.\cite{Beyer:2004sv, Alcubierre:2002kk}. For the lapse, it has the form
\begin{eqnarray}
	\partial_0\alpha=-\alpha^2h(\alpha,\phi,x^\mu)\Big[K-K_0(x^\mu)\Big]\, ,
\end{eqnarray}
which is named as ``hyperbolic K-driver" condition. For the shift, the ``hyperbolic Gamma driver" type condition 
\begin{eqnarray}
	\partial_0\beta^i&=&\alpha^2G(\alpha,\phi,x^\mu)B^i\, ,\\
	\partial_0B^i&=&e^{-4\phi}H(\alpha,\phi,x^\mu)\partial_0\tilde{\Gamma}^i-\eta(B^i,\alpha,x^\mu)
\end{eqnarray}
will be applied, where $G(\alpha,\phi,x^{\mu})$ and $H(\alpha,\phi,x^{\mu})$ are smooth, strictly positive functions, and $\eta(B^i,\alpha,x^{\mu})$ is a smooth function.

Freezing the coefficients in the differential equations at some fixed point and analyzing the linear constant coefficient problem by means of a Fourier transformation in space, we get
\begin{eqnarray}
	\partial_0\hat{\alpha}&=&-\alpha^2h\hat{K}+\text{l.o.}\, ,\label{BSSN_Fourier_alpha}\\
	\partial_0\hat{\beta}^i&=&\alpha^2G\hat{B}^i\, ,\\
	\partial_0\hat{B}^i&=&e^{-4\phi}H\Big[-\frac43\alpha m^{\prime}\tilde{\gamma}^{ij}(i\omega_j\hat{K})+2\alpha(m^{\prime}-1)\tilde{\gamma}^{ik}\tilde{\gamma}^{jl}(i\omega_j)\hat{\tilde{A}}_{kl}\nonumber\\
	&&-\tilde{\gamma}^{jk}\omega_j\omega_k\hat{\beta}^i-\frac13\tilde{\gamma}^{ij}\omega_j\omega_k\hat{\beta}^k+2\alpha m^{\prime}\frac{f^{\prime\prime}}{f^{\prime}}\tilde{\gamma}^{ij}(i\omega_j\hat{\psi})\Big]+\text{l.o.}\, ,\\
	\partial_0\hat{R}&=&\alpha\hat{\psi}\, ,\\
	\partial_0\hat{\psi}&=&-\alpha\gamma^{ij}\omega_i\omega_j\hat{R}+\text{l.o.}\, ,\\
	\partial_0\hat{\phi}&=&-\frac16\alpha\hat{K}+\frac{i}{6}\omega_k\hat{\beta}^k\, ,\\
	\partial_0\hat{\tilde{\gamma}}_{ij}&=&-2\alpha\hat{\tilde{A}}_{ij}+i\tilde{\gamma}_{ik}\omega_j\hat{\beta}^k+i\tilde{\gamma}_{jk}\omega_i\hat{\beta}^k-\frac{2i}{3}\tilde{\gamma}_{ij}\omega_k\hat{\beta}^k\, ,\\
	\partial_0\hat{K}&=&-\alpha\frac{f^{\prime\prime}}{f^{\prime}}\gamma^{ij}\omega_i\omega_j\hat{R}+\gamma^{ij}\omega_i\omega_j\hat{\alpha}+\text{l.o.}\, ,\\
	\partial_0\hat{\tilde{A}}_{ij}&=&\alpha e^{-4\phi}\Big[\frac12\tilde{\gamma}^{kl}\omega_k\omega_l\hat{\tilde{\gamma}}_{ij}+i\tilde{\gamma}_{k(i}\omega_{j)}\hat{\tilde{\Gamma}}^k+2\omega_i\omega_j\hat{\phi}+\omega_i\omega_j\frac{\hat{\alpha}}{\alpha}\Big]^{\text{TF}}\nonumber\\
	&&+\alpha e^{-4\phi}\frac{f^{\prime\prime}}{f^{\prime}}(\omega_i\omega_j\hat{R})^{\text{TF}}+\text{l.o.}\, ,\\
	\partial_0\hat{\tilde{\Gamma}}^i&=&-\frac43\alpha m\tilde{\gamma}^{ij}(i\omega_j\hat{K})+2\alpha(m-1)\tilde{\gamma}^{ik}\tilde{\gamma}^{jl}(i\omega_j\hat{\tilde{A}}_{kl})-\tilde{\gamma}^{jk}\omega_j\omega_k\hat{\beta}^i\nonumber\\
	&&-\frac13\tilde{\gamma}^{ij}\omega_j\omega_k\hat{\beta}^k+2\alpha m\frac{f^{\prime\prime}}{f^{\prime}}\tilde{\gamma}^{ij}(i\omega_j\hat{\psi})+\text{l.o.}\, ,\label{BSSN_Fourier_Gamma}
\end{eqnarray}
where a hat represents the Fourier transformation in space, for example, 
 $$\hat{\theta}(\omega)=\int\mathrm{d}^3x\theta(x)e^{-i\omega\cdot x}\, ,$$ 
 and l.o. denotes terms which depend on lower order spatial derivatives. 
 The parameter $m^{\prime}$ in the evolution equation for $B^i$ is allowed to be different from the parameter $m$~\cite{Beyer:2004sv}. 
 By writing
 $$\omega_i=|\omega|\tilde{\omega}_i\, ,\qquad |\omega|^2=\gamma^{ij}\omega_i\omega_j\, ,$$ 
 and introducing the variables
\begin{eqnarray}
	&&\hat{a}=i\alpha^{-1}|\omega|\hat{\alpha}\, ,\quad\hat{b}_i=i\alpha^{-1}|\omega|\gamma_{ij}\hat{\beta}^j\, ,\quad\hat{B}_i=\gamma_{ij}\hat{B}^j\label{B}\, ,\quad\hat{r}=i|\omega|\hat{R}\label{r}\, ,\nonumber\\
	&&\hat{\Phi}=i|\omega|\hat{\phi}\label{Phi}\, ,\quad\hat{l}_{ij}=i|\omega|e^{4\phi}\hat{\tilde{\gamma}}_{ij}\label{l}\, ,\quad\hat{L}_{ij}=e^{4\phi}\hat{\tilde{A}}_{ij}\,\quad\hat{\tilde{\Gamma}}_i=\tilde{\gamma}_{ij}\hat{\tilde{\Gamma}}^j\, ,
\end{eqnarray}
one can rewrite the system [Eq.(\ref{BSSN_Fourier_alpha})-Eq.(\ref{BSSN_Fourier_Gamma})] as a first order system. According to these variables, one gets a first order pseudodifferential system of the structure
\begin{eqnarray}
\label{Cauchy_problem}
	\partial_0\hat{u}=i|\omega|\alpha\mathbf{P}(\omega)\hat{u}+\text{l.o.}\, ,
\end{eqnarray}
where 
$$\hat{u}=\Big(\hat{a},\hat{b}_i,\hat{B}_i,\hat{r},\hat{\psi},\hat{\Phi},\hat{l}_{ij},\hat{K},\hat{L}_{ij},\hat{\tilde{\Gamma}}_i\Big)^{\text{T}}\, .$$
Since the shift cannot change a real eigenvalue into an imaginary one and it cannot affect the hyperbolicity of the system, the system (\ref{Cauchy_problem}) is strongly hyperbolic if and only if that $\mathbf{P}(\omega)$ is diagonalizable and has only real eigenvalues~\cite{Sarbach:2012pr,1989Initial}. 
An ingenious suggestion for doing these calculations is to decompose the eigenvalue equation 
\begin{eqnarray}
	\lambda\hat{u}=\mathbf{P}(\omega)\hat{u}
\end{eqnarray}
into orthogonal components with respect to $\tilde{\omega}_i$~\cite{Nagy:2004td,Sarbach:2019yso,Sarbach:2012pr}. Introduce the splitting 
\begin{eqnarray}
	X_{ij}&=&\tilde{\omega}_i\tilde{\omega}_jX+X^{\prime}q_{ij}/2+2\tilde{\omega}_{(i}X^{\prime}_{j)}+X^{\prime}_{\langle ij \rangle}\, ,\nonumber\\
	Y_i&=&Y^{\prime}_i+\tilde{\omega}_iY\, ,
\end{eqnarray}
where $q_{ij}=\gamma_{ij}-\tilde{\omega}_i\tilde{\omega}_j$ is the orthogonal projector to $\tilde{\omega}_i$, and
\begin{eqnarray}
	&&X=\tilde{\omega}^i\tilde{\omega}^jX_{ij}\, ,\quad X^{\prime}=q^{ij}X_{ij}\, ,\nonumber\\
	&& X^{\prime}_{i}=q_i{}^k\tilde{\omega}^lX_{kl}\, ,\quad X^{\prime}_{\langle ij\rangle}=q_i{}^kq_j{}^l\big(X_{kl}-X^{\prime}q_{kl}/2\big)\, ,\nonumber\\
	&&Y=\tilde{\omega}_iY^i\, ,\quad Y^{\prime}_i=q_i{}^jY_j\, .
\end{eqnarray}
In this subsection, $X_{ij}$ is choosen to be $\hat{l}_{ij}$, $\hat{L}_{ij}$, and $Y_i$ is choosen to be $\hat{b}_i$, $\hat{B}_i$, $\hat{\tilde{\Gamma}}_i$. Hence, $\mathbf{P}(\omega)$ can be decomposed into three independent parts, and it is written as
\begin{eqnarray}
	\mathbf{P}=\left[\begin{matrix}
		\mathbf{P}^S & \mathbf{0} & \mathbf{0}\\
		\mathbf{0} & \mathbf{P}^V & \mathbf{0}\\
		\mathbf{0} & \mathbf{0} & \mathbf{P}^T
	\end{matrix}\right]\, ,
\end{eqnarray} 
where $\mathbf{P}^S,\mathbf{P}^V,\mathbf{P}^T$ denote scalar part, vector part, and tensor part, respectively. After some caculations, the results are showed as follows,
\begin{eqnarray}
	\mathbf{P}^S\begin{bmatrix}
		\hat{a}\\
		\hat{b}\\
		\hat{B}\\
		\hat{r}\\
		\hat{\psi}\\
		\hat{\Phi}\\
		\hat{l}\\
		\hat{l}^{\prime}\\
		\hat{K}\\
		\hat{L}\\
		\hat{\tilde{\Gamma}}
	\end{bmatrix}
	=\begin{bmatrix}
		0 & 0 & 0 & 0 & 0 & 0 & 0 & 0 & -h & 0 & 0\\
		0 & 0 & G & 0 & 0 & 0 & 0 & 0 & 0 & 0 & 0\\
		0 & \frac43H & 0 & 0 & 2m^{\prime}\frac{f^{\prime\prime}}{f^{\prime}}H & 0 & 0 & 0 & -\frac43m^{\prime}H & 2(m^{\prime}-1)H & 0\\
		0 & 0 & 0 & 0 & 1 & 0 & 0 & 0 & 0 & 0 & 0\\
		0 & 0 & 0 & 1 & 0 & 0 & 0 & 0 & 0 & 0 & 0\\
		0 & \frac16 & 0 & 0 & 0 & 0 & 0 & 0 & -\frac16 & 0 & 0\\
		0 & \frac43 & 0 & 0 & 0 & 0 & 0 & 0 & 0 & -2 & 0\\
		0 & -\frac43 & 0 & 0 & 0 & 0 & 0 & 0 & 0 & 2 & 0\\
		-1 & 0 & 0 & \frac{f^{\prime\prime}}{f^{\prime}} & 0 & 0 & 0 & 0 & 0 & 0 & 0\\
		-\frac23 & 0 & 0 & -\frac{2f^{\prime\prime}}{3f^{\prime}} & 0 & -\frac43 & -\frac13 & \frac16 & 0 & 0 & \frac23\\
		0 & \frac43 & 0 & 0 & 2m\frac{f^{\prime\prime}}{f^{\prime}} & 0 & 0 & 0 & -\frac43m & 2(m-1) & 0
	\end{bmatrix}\begin{bmatrix}
	\hat{a}\\
	\hat{b}\\
	\hat{B}\\
	\hat{r}\\
	\hat{\psi}\\
	\hat{\Phi}\\
	\hat{l}\\
	\hat{l}^{\prime}\\
	\hat{K}\\
	\hat{L}\\
	\hat{\tilde{\Gamma}}
\end{bmatrix}\, ,
\end{eqnarray}
\begin{eqnarray}
	\mathbf{P}^V\begin{bmatrix}
		\hat{b}^{\prime}_i\\
		\hat{B}^{\prime}_i\\
		\hat{l}^{\prime}_i\\
		\hat{L}^{\prime}_i\\
		\hat{\tilde{\Gamma}}^{\prime}_i
	\end{bmatrix}=\begin{bmatrix}
	0 & G & 0 & 0 & 0\\
	H & 0 & 0 & 2(m^{\prime}-1)H & 0\\
	1 & 0 & 0 & -2 & 0\\
	0 & 0 & -\frac12 & 0 & \frac12\\
	1 & 0 & 0 & 2(m-1) & 0
\end{bmatrix}\begin{bmatrix}
\hat{b}^{\prime}_i\\
\hat{B}^{\prime}_i\\
\hat{l}^{\prime}_i\\
\hat{L}^{\prime}_i\\
\hat{\tilde{\Gamma}}^{\prime}_i
\end{bmatrix}\, ,
\end{eqnarray}
and
\begin{eqnarray}
\quad\mathbf{P}^T\begin{bmatrix}
	\hat{l}^{\prime}_{\langle ij\rangle}\\
	\hat{L}^{\prime}_{\langle ij\rangle}
\end{bmatrix}=\begin{bmatrix}
0 & -2\\
-\frac12 & 0
\end{bmatrix}\begin{bmatrix}
\hat{l}^{\prime}_{\langle ij\rangle}\\
\hat{L}^{\prime}_{\langle ij\rangle}
\end{bmatrix}\, .
\end{eqnarray}
The eigenvalues of the matrix $\mathbf{P}^S$ are $0\, ,\pm 1\, ,\pm\sqrt{h}\, ,\pm\sqrt{4GH/3}\, ,\pm\sqrt{(4m-1)/3}$, where $0$ is the triple root. 
The eigenvalues of the matrix $\mathbf{P}^V$ are $0\, ,\pm\sqrt{GH}\, ,\pm\sqrt{m}$. The eigenvalues of the matrix $\mathbf{P}^T$ are $\pm 1$. Therefore,  to guarantee the weak hyperbolicity, we have to set $h>0$, $GH>0$, $m>1/4$. 

Furthermore, provided that $m^{\prime}=1$, the matrix $\mathbf{P}$ is diagonalizable only if 
\begin{eqnarray}\label{BSSN_condition_1}
	h\neq1\, ,\quad GH\neq3/4\, ,\quad h\neq4GH/3\, .
\end{eqnarray} 
Provided that $m^{\prime}\neq1$, the matrix $\mathbf{P}$ is diagonalizable only if 
\begin{eqnarray}\label{BSSN_condition_2}
	h\neq1\, ,\quad GH\neq3/4\, ,\quad h\neq4GH/3\, ,\quad 4GH\neq4m-1\, ,\quad m\neq GH\, .
\end{eqnarray} 
These are the conditions for the strong hyperbolicity.

\section{ADM formulation with modified harmonic gauge}\label{section5}

Since the ADM equations are not strongly hyperbolic with a fixed  shift $\beta^i$ and a dynamical lapse $\alpha$ whose evolution are denoted by a memeber of the Bona-Masso family~\cite{Mongwane:2016qtz}, in this time, we consider the case where $\beta$ and $\alpha$ are both dynamic variables. This consideration of gauge condition called the modified harmonic gauge is different from the one by Bona-Masso \cite{Bona:1994dr}. The harmonic gauge of Einstein's field equations has many generalizations. One of them is to add a given source functions, denoted by $H^{\nu}$,  
into the usual harmonic condition~\cite{Friedrich:1996hq,Garfinkle:2001ni}, and the gauge is written as
\begin{eqnarray}
	\nabla^\mu\nabla_\mu x^\nu=H^\nu\, .
\end{eqnarray}
For keeping general covariance, the generalized harmonic gauge condition can be expressed as~\cite{Hawking:1973uf}
\begin{eqnarray}\label{harmonicH}
	g^{\alpha\beta}\Big({}^4\Gamma^{\mu}{}_{\alpha\beta}-{}^4\mathring{\Gamma}^{\mu}{}_{\alpha\beta}\Big)+H^\mu=0\, ,
\end{eqnarray}
where the Christoffel symbols ${}^4\mathring{\Gamma}^\mu{}_{\alpha\beta}$ come from a fixed smooth background metric $\mathring{g}_{\alpha\beta}$.
Assuming that the background metric $\mathring{g}_{\alpha\beta}$ is Minkowski in Cartesian coordinates for simplicity. This means ${}^4\mathring{\Gamma}^\mu{}_{\alpha\beta}$ is vanished. Therefore, in this subsection, we choose a modified harmonic formulation given by
\begin{eqnarray}\label{modified_harmonicH}
	\tilde{g}^{\alpha\beta}{}^4\Gamma^\mu{}_{\alpha\beta}+H^\mu=0\, ,
\end{eqnarray}
where $\tilde{g}^{\alpha\beta}$ is defined as $$\tilde{g}^{\alpha\beta}\equiv g^{\alpha\beta}+h^{\alpha\beta}\, .$$ The modified quanties $h^{\alpha\beta}$ satisfy 
\begin{eqnarray}
	h^{\alpha\beta}{}^4\Gamma^t{}_{\alpha\beta}&=&\frac{1-F}{\alpha}K\, ,\\
	h^{\alpha\beta}{}^4\Gamma^i{}_{\alpha\beta}&=&\frac{1-p}{\alpha}\gamma^{ij}\partial_j\alpha+(p-1)\gamma^{ij}\gamma^{kl}\Big(\partial_k\gamma_{jl}-\frac12\partial_j\gamma_{kl}\Big)+\frac{F-1}{\alpha}\beta^iK\, ,
\end{eqnarray}
where ${}^4\Gamma^t{}_{\alpha\beta}$ and $K$ are obtained from the original metric~(\ref{metric}) and $F$, $p$ are constants. 
Note that when $F=1$ and $p=1$, $h^{\alpha\beta}=0$, Eq.(\ref{modified_harmonicH}) becomes Eq.(\ref{harmonicH}). We will show when the following conditions
\begin{eqnarray}\label{condition_ADM_harmonic}
	F\neq p\, ,\quad F\neq1\, ,\quad p\neq1\, ,\quad F>0\, ,\quad p>0\, 
\end{eqnarray}
are true, the evolution equation~[Eq.(\ref{ADM_R})-Eq.(\ref{ADM_Kij})] with modified harmonic condition (\ref{modified_harmonicH}) leads a well-posed formulation. Due to the fact that this system is a mixed first/second order system, with same ideas as previous sections, first order pseudodifferential reduction is used. We obtain 
\begin{eqnarray}
	\partial_0\hat{\alpha}&=&-\alpha^2F\gamma^{ij}\hat{K}_{ij}+\text{l.o.}\, ,\\
	\partial_0\hat{\beta}^i&=&-\alpha p\gamma^{ij}(i\omega_j\hat{\alpha})+\alpha^2p\gamma^{ij}\gamma^{kl}(i\omega_k\hat{\gamma}_{jl}-\frac{i}{2}\omega_j\hat{\gamma}_{kl})+\text{l.o.}\, ,\\
	\partial_0\hat{R}&=&\alpha\hat{\psi}\, ,\\
	\partial_0\hat{\psi}&=&-\alpha\gamma^{ij}\omega_i\omega_j\hat{R}+\text{l.o.}\, ,\\
	\partial_0\hat{\gamma}_{ij}&=&-2\alpha \hat{K}_{ij}+\gamma_{jk}(i\omega_i\hat{\beta}^k)+\gamma_{ik}(i\omega_j\hat{\beta}^k)\, ,\\
	\partial_0\hat{K}_{ij}&=&\omega_i\omega_j\hat{\alpha}+\frac{\alpha}{2}\gamma^{kl}(\omega_k\omega_l\hat{\gamma}_{ij}+\omega_i\omega_j\hat{\gamma}_{kl}-\omega_i\omega_k\hat{\gamma}_{lj}-\omega_j\omega_k\hat{\gamma}_{li})+\alpha\frac{f^{\prime\prime}}{f^{\prime}}\omega_i\omega_j\hat{R}+\text{l.o.}\, .
\end{eqnarray}	
After introducing the variables 
\begin{eqnarray}
	\hat{a}=i\alpha^{-1}|\omega|\hat{\alpha}\, ,\quad\hat{b}_i=i\alpha^{-1}|\omega|\gamma_{ij}\hat{\beta}^j\, ,\quad\hat{r}=i|\omega|\hat{R}\, ,\quad\hat{l}_{ij}=i|\omega|\hat{\gamma}_{ij}\, ,
\end{eqnarray}
and the splitting
\begin{eqnarray}
	\hat{l}_{ij}&=&\tilde{\omega}_i\tilde{\omega}_j\hat{l}+\hat{l}^{\prime}\frac{q_{ij}}{2}+2\tilde{\omega}_{(i}\hat{l}^{\prime}_{j)}+\hat{l}^{\prime}_{\langle ij\rangle}\, ,\nonumber\\
	\hat{K}_{ij}&=&\tilde{\omega}_i\tilde{\omega}_j\hat{\mathbb{K}}+\hat{\mathbb{K}}^{\prime}\frac{q_{ij}}{2}+2\tilde{\omega}_{(i}\hat{\mathbb{K}}^{\prime}_{j)}+\hat{\mathbb{K}}^{\prime}_{\langle ij\rangle}\, ,\nonumber\\
	\hat{b}_i&=&\hat{b}^{\prime}_i+\tilde{\omega}_i\hat{b}\, ,
\end{eqnarray}
we have the following results,
\begin{eqnarray}\label{harmonic_PS}
	\mathbf{P}^S\begin{bmatrix}
		\hat{a}\\
		\hat{b}\\
		\hat{r}\\
		\hat{\psi}\\
		\hat{l}\\
		\hat{l}^{\prime}\\
		\hat{\mathbb{K}}\\
		\hat{\mathbb{K}}^{\prime}
	\end{bmatrix}=\begin{bmatrix}
	0 & 0 & 0 & 0 & 0 & 0 & -F & -F\\
	-p & 0 & 0 & 0 & \frac{p}{2} & -\frac{p}{2} & 0 & 0\\
	0 & 0 & 0 & 1 & 0 & 0 & 0 & 0\\
	0 & 0 & 1 & 0 & 0 & 0 & 0 & 0\\
	0 & 2 & 0 & 0 & 0 & 0 & -2 & 0\\
	0 & 0 & 0 & 0 & 0 & 0 & 0 & -2\\
	-1 & 0 & -\frac{f^{\prime\prime}}{f^{\prime}} & 0 & 0 & -\frac12 & 0 & 0\\
	0 & 0 & 0 & 0 & 0 & -\frac12 & 0 & 0\\
\end{bmatrix}\begin{bmatrix}
\hat{a}\\
\hat{b}\\
\hat{r}\\
\hat{\psi}\\
\hat{l}\\
\hat{l}^{\prime}\\
\hat{\mathbb{K}}\\
\hat{\mathbb{K}}^{\prime}
\end{bmatrix}\, ,
\end{eqnarray}

\begin{eqnarray}\label{harmonic_PV}
	\mathbf{P}^V\begin{bmatrix}
		\hat{b}^{\prime}_i\\
		\hat{l}^{\prime}_i\\
		\hat{\mathbb{K}}^{\prime}_i
	\end{bmatrix}=\begin{bmatrix}
	0 & p & 0\\
	1 & 0 & -2\\
	0 & 0 & 0
\end{bmatrix}\begin{bmatrix}
\hat{b}^{\prime}_i\\
\hat{l}^{\prime}_i\\
\hat{\mathbb{K}}^{\prime}_i
\end{bmatrix}\, ,
\end{eqnarray}
and
\begin{eqnarray}\label{harmonic_PT}
	\mathbf{P}^T\begin{bmatrix}
		\hat{l}^{\prime}_{\langle ij\rangle}\\
		\hat{\mathbb{K}}^{\prime}_{\langle ij\rangle}
	\end{bmatrix}=\begin{bmatrix}
	0 & -2\\
	-\frac12 & 0
\end{bmatrix}\begin{bmatrix}
\hat{l}^{\prime}_{\langle ij\rangle}\\
\hat{\mathbb{K}}^{\prime}_{\langle ij\rangle}
\end{bmatrix}\, .
\end{eqnarray}
The eigenvalues of the matrix $\mathbf{P}^S$ are $\pm1\, ,\pm\sqrt{F}\, ,\pm\sqrt{p}$, where $\pm1$ are the double root. The eigenvalues of the matrix $\mathbf{P}^V$ are $\pm\sqrt{p}\, ,0$. The eigenvalues of the matrix $\mathbf{P}^T$ are $\pm1$. 

Hence, $F>0$ and $p>0$ guarantee the weak hyperbolicity of the evolution equation [Eq.(\ref{ADM_R})-Eq.(\ref{ADM_Kij})]. Furthermore, conditions (\ref{condition_ADM_harmonic}) means the evolution equations [Eq.(\ref{ADM_R})-Eq.(\ref{ADM_Kij})] with modified harmonic donditions (\ref{modified_harmonicH}) are strong hyperbolic.

\section{modified Z4 formulation with modified harmonic gauge}\label{section4}

We extend equations of motion (\ref{EOM}) in a general covariant way by introducing an extra four-vector $\mathcal{Z}^a$~\cite{Bona:2002fq,Bona:2004yp,Gundlach:2005eh}, so that the set of basic fields will become $\left\lbrace g_{\mu\nu}, \mathcal{Z}_\mu\right\rbrace$. To be specific, the modification is carried out in the following way,
\begin{eqnarray}\label{EOM_Z4}
	\Sigma_{ab}-\kappa^2T_{ab}=0\to \Sigma_{ab}-\kappa^2T_{ab}+l_1\nabla_a\mathcal{Z}_b+l_2\nabla_b\mathcal{Z}_a-l_3g_{ab}\nabla^c\mathcal{Z}_c-k_1\Big(n_a\mathcal{Z}_b+n_b\mathcal{Z}_a+k_2n^c\mathcal{Z}_cg_{ab}\Big)=0\, ,
\end{eqnarray}
where $k_1$ and $k_2$ are real constants. In the above equations, we have added three other different parameters 
$$l_1\, ,\quad l_2\, ,\quad l_3 \, ,\qquad l_1+l_2-l_3\neq0 $$ 
into the uaual Z4 formulation. These three parameters can be unequal with each other. Note that it is a key point for the strong hyperbolicity. Splitting the four-vector $\mathcal{Z}^a$ as $\mathcal{Z}^a=Z^a+n^a\Theta$ with $Z^a=\gamma^a{}_b\mathcal{Z}^b$ and $\Theta=-n^a\mathcal{Z}_a$.

The harmonic gauge condition (\ref{harmonicH}) in this subsection is modified as the following form~\cite{Bona:2004yp}
\begin{eqnarray}
	\partial_0\alpha&=&-\alpha^2\zeta(K-m\Theta)\, ,\label{harmonic_Z4_alpha}\\
	\partial_0\beta^i&=&-\alpha^2(2\mu V^i+c\partial^i\ln\alpha-d\partial^i\ln\sqrt{\gamma})-\xi\beta^i\, ,\label{harmonic_Z4_beta}
\end{eqnarray} 
where
\begin{eqnarray}
	V_i=\partial_i\ln\sqrt{\gamma}-\frac12\partial^j\gamma_{ji}-Z_i\, .\label{harmonic_Z4_V}
\end{eqnarray}
What is worth mentioning is that when 
\begin{eqnarray}
	\zeta=1\, ,\quad m=0\, ,\quad\mu=1\, ,\quad c=1\, ,\quad d=1\, ,\quad\xi=0\, ,\quad Z_i=0\, ,
\end{eqnarray}
Eq.(\ref{harmonic_Z4_alpha}) and Eq.(\ref{harmonic_Z4_beta}) are going to be Eq.(\ref{harmonicH}) with $H^{\mu}=0$. Projecting Eq.(\ref{EOM_Z4}) onto $n^i$ or $\gamma_{ij}$ with some caculations, we finally arrive to the evolution system
\begin{eqnarray}
	\partial_0\Theta&=&\frac{\alpha}{l_1+l_2-l_3}\Big[\frac12f-\frac12Rf^{\prime}+\frac12\Big(\mathcal{R}+K^2-K_{ij}K^{ij}\Big)f^{\prime}-\Big(D^iD_iR+K\psi\Big)f^{\prime\prime}-f^{\prime\prime\prime}D^iRD_iR-\kappa^2\rho\Big]\label{Z4_Theta}\nonumber\\
	&&-Z^kD_k\alpha+\frac{\alpha l_3}{l_1+l_2-l_3}D_kZ^k-\frac{\alpha l_3}{l_1+l_2-l_3}\Theta K-\frac{\alpha}{l_1+l_2-l_3}k_1(2+k_2)\Theta\, ,
\end{eqnarray}
\begin{eqnarray}
	\partial_0Z_i&=&\frac{\alpha}{l_2}\Big[f^{\prime}\Big(D_jK_i{}^j-D_iK\Big)+f^{\prime\prime}\Big(K_i{}^jD_jR+D_i\psi\Big)+f^{\prime\prime\prime}\psi D_iR-\kappa^2S_i\Big]\nonumber\\
	&&+\frac{\alpha l_1}{l_2}D_i\Theta-\Theta D_i\alpha-\frac{\alpha(l_1+l_2)}{l_2}K_{ij}Z^j-\frac{\alpha k_1}{l_2}Z_i+Z_k\partial_i\beta^k\, ,
\end{eqnarray}
\begin{eqnarray}
	\partial_0R&=&\alpha\psi\, ,
\end{eqnarray}
\begin{eqnarray}
	\partial_0\psi&=&\frac{\alpha}{3f^{\prime\prime}}\Biggl\{-\frac{3(l_1+l_2)}{2(l_1+l_2-l_3)}f+\frac{l_1+l_2+2l_3}{2(l_1+l_2-l_3)}Rf^{\prime}+\frac{2(l_1+l_2)+l_3}{l_1+l_2-l_3}\Big(D_iD^iR+K\psi\Big)f^{\prime\prime}+3a^iD_iRf^{\prime\prime}\nonumber\\
	&&+\frac{2(l_1+l_2)+l_3}{l_1+l_2-l_3}D^iRD_iRf^{\prime\prime\prime}-3\psi^2f^{\prime\prime\prime}-\kappa^2S+\frac{3l_3}{l_1+l_2-l_3}\kappa^2\rho+\frac{(l_1+l_2-4l_3)(l_1+l_2)}{l_1+l_2-l_3}D^iZ_i\nonumber\\
	&&+\frac{l_1+l_2-4l_3}{2(l_1+l_2-l_3)}\Big(\mathcal{R}+K^2-K_{ij}K^{ij}\Big)f^{\prime}-\frac{(l_1+l_2-4l_3)(l_1+l_2)}{l_1+l_2-l_3}\Theta K\nonumber\\
	&&+\Big[-\frac{l_1+l_2-4l_3}{l_1+l_2-l_3}k_1(2+k_2)+2k_1(1+2k_2)\Big]\Theta\Biggr\}\, ,
\end{eqnarray}
\begin{eqnarray}
	\partial_0\gamma_{ij}&=&-2\alpha K_{ij}+\gamma_{ik}\partial_j\beta^k+\gamma_{jk}\partial_i\beta^k\, ,
\end{eqnarray}
and
\begin{eqnarray}
	\partial_0K_{ij}&=&\alpha\Big(\mathcal{R}_{ij}-2K_i{}^kK_{jk}+KK_{ij}-\frac{1}{\alpha}D_iD_j\alpha\Big)-\alpha\frac{f^{\prime\prime}}{f^{\prime}}\Big(D_iD_jR+\psi K_{ij}\Big)-\alpha\frac{f^{\prime\prime\prime}}{f^{\prime}}D_iRD_jR\nonumber\\
	&&+\frac{\alpha}{f^{\prime}}\Big[-\frac16Rf^{\prime}+\frac13f^{\prime\prime\prime}D^kRD_kR+\frac13\Big(D^kD_kR+K\psi\Big)f^{\prime\prime}-\frac16\Big(\mathcal{R}+K^2-K_{kl}K^{kl}\Big)f^{\prime}\nonumber\\
	&&-\frac{l_1+l_2}{3}D^kZ_k+\frac{l_1+l_2}{3}\Theta K+\frac13\kappa^2S\Big]\gamma_{ij}+2K_{k(i}\partial_{j)}\beta^k\label{Z4_Kij}\nonumber\\
	&&+\frac{\alpha}{f^{\prime}}\Big[-\kappa^2S_{ij}+(l_1+l_2)D_{(i}Z_{j)}-(l_1+l_2)\Theta K_{ij}\Big]\, .
\end{eqnarray}
Only remaining the princinpal terms and doing the Fourier transformation, we get
\begin{eqnarray}
	\partial_0\hat{\alpha}&=&-\alpha^2\zeta(\gamma^{ij}\hat{K}_{ij}-m\hat{\Theta})\, ,\\
	\partial_0\hat{\beta}_i&=&-\alpha^2\Big[2\mu\Big(\frac12\gamma^{kl}i\omega_i\hat{\gamma}_{kl}-\frac12i\omega^j\hat{\gamma}_{ji}-\hat{Z}_i\Big)-\frac{d}{2}\gamma^{kl}i\omega_i\hat{\gamma}_{kl}\Big]-\alpha ci\omega_i\hat{\alpha}+\text{l.o.}\, ,\\
	\partial_0\hat{\Theta}&=&\frac{\alpha}{4(l_1+l_2-l_3)}f^{\prime}\gamma^{ij}\gamma^{kl}\Big[(i\omega_i)(i\omega_l)\hat{\gamma}_{kj}+(i\omega_k)(i\omega_j)\hat{\gamma}_{il}-(i\omega_i)(i\omega_j)\hat{\gamma}_{kl}-(i\omega_k)(i\omega_l)\hat{\gamma}_{ij}\Big]\nonumber\\
	&&-\frac{\alpha}{l_1+l_2-l_3}f^{\prime\prime}\gamma^{ij}(i\omega_i)(i\omega_j)\hat{R}+\frac{\alpha l_3}{l_1+l_2-l_3}\gamma^{ij}i\omega_i\hat{Z}_j+\text{l.o.}\, ,\\
	\partial_0\hat{Z}_i&=&\frac{\alpha}{l_2}f^{\prime}\Big(\gamma^{jk}i\omega_k\hat{K}_{ij}-\gamma^{jk}i\omega_i\hat{K}_{jk}\Big)+\frac{\alpha}{l_2}f^{\prime\prime}i\omega_i\hat{\psi}+\frac{\alpha l_1}{l_2}i\omega_i\hat{\Theta}+\text{l.o.}\, ,\\
	\partial_0\hat{R}&=&\alpha\hat{\psi}\, ,\\
	\partial_0\hat{\psi}&=&\frac{\alpha[2(l_1+l_2)+l_3]}{3(l_1+l_2-l_3)}\gamma^{ij}(i\omega_i)(i\omega_j)\hat{R}+\frac{\alpha(l_1+l_2-4l_3)(l_1+l_2)}{3f^{\prime\prime}(l_1+l_2-l_3)}\gamma^{ij}i\omega_i\hat{Z}_j\nonumber\\
	&&+\frac{\alpha(l_1+l_2-4l_3)f^{\prime}}{12f^{\prime\prime}(l_1+l_2-l_3)}\gamma^{ij}\gamma^{kl}\Big[(i\omega_i)(i\omega_l)\hat{\gamma}_{kj}+(i\omega_k)(i\omega_j)\hat{\gamma}_{il}-(i\omega_i)(i\omega_j)\hat{\gamma}_{kl}-(i\omega_k)(i\omega_l)\hat{\gamma}_{ij}\Big]+\text{l.o.}\, ,
\end{eqnarray}
\begin{eqnarray}
	\partial_0\hat{\gamma}_{ij}&=&-2\alpha \hat{K}_{ij}+\gamma_{jk}(i\omega_i\hat{\beta}^k)+\gamma_{ik}(i\omega_j\hat{\beta}^k)\, ,
\end{eqnarray}
and
\begin{eqnarray}	
	\partial_0\hat{K}_{ij}&=&\frac{\alpha}{2}\gamma^{kl}\Big[(i\omega_i)(i\omega_l)\hat{\gamma}_{kj}+(i\omega_k)(i\omega_j)\hat{\gamma}_{il}-(i\omega_i)(i\omega_j)\hat{\gamma}_{kl}-(i\omega_k)(i\omega_l)\hat{\gamma}_{ij}\Big]-(i\omega_i)(i\omega_j)\hat{\alpha}\nonumber\\
	&&-\alpha\frac{f^{\prime\prime}}{f^{\prime}}(i\omega_i)(i\omega_j)\hat{R}+\gamma_{ij}\Biggl\{\frac{\alpha f^{\prime\prime}}{3f^{\prime}}\gamma^{kl}(i\omega_k)(i\omega_l)\hat{R}-\frac{\alpha}{12}\gamma^{mn}\gamma^{kl}\Big[(i\omega_m)(i\omega_l)\hat{\gamma}_{kn}+(i\omega_k)(i\omega_n)\hat{\gamma}_{ml}\nonumber\\
	&&-(i\omega_m)(i\omega_n)\hat{\gamma}_{kl}-(i\omega_k)(i\omega_l)\hat{\gamma}_{mn}\Big]-\frac{\alpha(l_1+l_2)}{3f^{\prime}}\gamma^{kl}i\omega_k\hat{Z}_l\Biggr\}+\frac{\alpha(l_1+l_2)}{2f^{\prime}}\Big(i\omega_i\hat{Z}_j+i\omega_j\hat{Z}_i\Big)+\text{l.o.}\, .
\end{eqnarray}
After introducing the variables
\begin{eqnarray}
	\hat{a}=i\alpha^{-1}|\omega|\hat{\alpha}\, ,\quad\hat{b}_i=i\alpha^{-1}|\omega|\gamma_{ij}\hat{\beta}^j\, ,\quad\hat{r}=i|\omega|\hat{R}\, ,\quad\hat{l}_{ij}=i|\omega|\hat{\gamma}_{ij}\, ,
\end{eqnarray}
and the splitting
\begin{eqnarray}
	\hat{l}_{ij}&=&\tilde{\omega}_i\tilde{\omega}_j\hat{l}+\hat{l}^{\prime}\frac{q_{ij}}{2}+2\tilde{\omega}_{(i}\hat{l}^{\prime}_{j)}+\hat{l}^{\prime}_{\langle ij\rangle}\, ,\nonumber\\
	\hat{K}_{ij}&=&\tilde{\omega}_i\tilde{\omega}_j\hat{\mathbb{K}}+\hat{\mathbb{K}}^{\prime}\frac{q_{ij}}{2}+2\tilde{\omega}_{(i}\hat{\mathbb{K}}^{\prime}_{j)}+\hat{\mathbb{K}}^{\prime}_{\langle ij\rangle}\, ,\nonumber\\
	\hat{b}_i&=&\hat{b}^{\prime}_i+\tilde{\omega}_i\hat{b}\, ,\nonumber\\
	\hat{Z}_i&=&\hat{Z}^{\prime}_i+\tilde{\omega}_i\hat{Z}\, ,
\end{eqnarray}
we have the following results,
\begin{eqnarray}
	\mathbf{P}^S\begin{bmatrix}
		\hat{a}\\
		\hat{b}\\
		\hat{\Theta}\\
		\hat{Z}\\
		\hat{r}\\
		\hat{\psi}\\
		\hat{l}\\
		\hat{l}^{\prime}\\
		\hat{\mathbb{K}}\\
		\hat{\mathbb{K}}^{\prime}
	\end{bmatrix}=\begin{bmatrix}
	0 & 0 & m\zeta & 0 & 0 & 0 & 0 & 0 & -\zeta & -\zeta\\
	-c & 0 & 0 & 2\mu & 0 & 0 & \frac{d}{2} & \frac{d}{2}-\mu & 0 & 0\\
	0 & 0 & 0 & \frac{l_3}{l_1+l_2-l_3} & -\frac{f^{\prime\prime}}{l_1+l_2-l_3} & 0 & 0 & -\frac{f^{\prime}}{2(l_1+l_2-l_3)} & 0 & 0\\
	0 & 0 & \frac{l_1}{l_2} & 0 & 0 & \frac{f^{\prime\prime}}{l_2} & 0 & 0 & 0 & -\frac{f^{\prime}}{l_2}\\
	0 & 0 & 0 & 0 & 0 & 1 & 0 & 0 & 0 & 0\\
	0 & 0 & 0 & \frac{(l_1+l_2-4l_3)(l_1+l_2)}{3f^{\prime\prime}(l_1+l_2-l_3)} & \frac{2(l_1+l_2)+l_3}{3(l_1+l_2-l_3)} & 0 & 0 & -\frac{(l_1+l_2-4l_3)f^{\prime}}{6f^{\prime\prime}(l_1+l_2-l_3)} & 0 & 0\\
	0 & 2 & 0 & 0 & 0 & 0 & 0 & 0 & -2 & 0\\
	0 & 0 & 0 & 0 & 0 & 0 & 0 & 0 & 0 & -2\\
	-1 & 0 & 0 & \frac{2(l_1+l_2)}{3f^{\prime}} & -\frac{2f^{\prime\prime}}{3f^{\prime}} & 0 & 0 & -\frac13 & 0 & 0\\
	0 & 0 & 0 & -\frac{2(l_1+l_2)}{3f^{\prime}} & \frac{2f^{\prime\prime}}{3f^{\prime}} & 0 & 0 & -\frac16 & 0 & 0
\end{bmatrix}\begin{bmatrix}
\hat{a}\\
\hat{b}\\
\hat{\Theta}\\
\hat{Z}\\
\hat{r}\\
\hat{\psi}\\
\hat{l}\\
\hat{l}^{\prime}\\
\hat{\mathbb{K}}\\
\hat{\mathbb{K}}^{\prime}
\end{bmatrix}\, ,
\end{eqnarray}
\begin{eqnarray}
	\mathbf{P}^V\begin{bmatrix}
		\hat{b}^{\prime}_i\\
		\hat{Z}^{\prime}_i\\
		\hat{l}^{\prime}_i\\
		\hat{\mathbb{K}}^{\prime}_i
	\end{bmatrix}=\begin{bmatrix}
	0 & 2\mu & \mu & 0\\
	0 & 0 & 0 & \frac{f^{\prime}}{l_2}\\
	1 & 0 & 0 & -2\\
	0 & \frac{l_1+l_2}{2f^{\prime}} & 0 & 0
\end{bmatrix}\begin{bmatrix}
\hat{b}^{\prime}_i\\
\hat{Z}^{\prime}_i\\
\hat{l}^{\prime}_i\\
\hat{\mathbb{K}}^{\prime}_i
\end{bmatrix}\, ,
\end{eqnarray}
and
\begin{eqnarray}
	\mathbf{P}^T\begin{bmatrix}
		\hat{l}^{\prime}_{\langle ij\rangle}\\
		\hat{\mathbb{K}}^{\prime}_{\langle ij\rangle}
	\end{bmatrix}=\begin{bmatrix}
		0 & -2\\
		-\frac12 & 0
	\end{bmatrix}\begin{bmatrix}
		\hat{l}^{\prime}_{\langle ij\rangle}\\
		\hat{\mathbb{K}}^{\prime}_{\langle ij\rangle}
	\end{bmatrix}\, .
\end{eqnarray}
The eigenvalues of the matrix $\mathbf{P}^S$ are $\pm\sqrt{l_1/l_2}\, ,\pm1\, ,\pm\sqrt{\zeta}\, ,\pm\sqrt{d}$, where $\pm1$ are the double root. The eigenvalues of the matrix $\mathbf{P}^V$ are $\pm\sqrt{(l_1+l_2)/(2l_2)}\, ,\pm\sqrt{\mu}$. The eigenvalues of the matrix $\mathbf{P}^T$ are $\pm1$. 

Hence, $l_1l_2>0$, $\zeta>0$, $d>0$ and $\mu>0$ guarantee the weak hyperbolicity of the evolution equation [Eq.(\ref{Z4_Theta})-Eq.(\ref{Z4_Kij})]. Furthermore, the condition of strong hyperbolicity for the evolution equation [Eq.(\ref{Z4_Theta})-Eq.(\ref{Z4_Kij})] with modified harmonic conditions [Eq.(\ref{harmonic_Z4_alpha})-Eq.(\ref{harmonic_Z4_V})] is given by
\begin{eqnarray}
	&&l_1\neq l_2\, ,\quad l_1\neq\zeta l_2\, ,\quad l_1\neq dl_2\, ,\quad\zeta\neq1\, ,\nonumber\\
	&&d\neq1\, ,\quad\zeta\neq d\, ,\quad l_1+l_2\neq2l_2\mu\, .
\end{eqnarray}

\section{ conclusions and discussion}\label{section6}
In this paper, without using the equivalence between $f(R)$ gravity and Brans-Dicke theory, the initial value problem (IVP) of $f(R)$ gravity has been systematically studied. Three formulations have been considered. 
All of them are first order in time and second order in space, and are  based on the ADM decomposition of theory. It is found that these formulations are all strongly hyperbolic with  suitable gauge conditions. 

The first order pseudodifferential reduction performed in the space derivatives is the main tool used to analyse the hyperbolicity of these three formulations. There are no new constraints added to the system since this technique does not increase the number of equations. It emphasizes that well-posedness essentially captures the absence of divergent behavior in the high frequency limit of the solutions for a given system~\cite{Nagy:2004td}.

For the BSSN formulation with the so-called ``hyperbolic $K$-driver" condition and the ``hyperbolic Gamma driver" condition, the condition to keep the strong hyperbolicity is given by Eq.(\ref{BSSN_condition_1}) and Eq.(\ref{BSSN_condition_2}). For general relativity, with the same gauge condition, one can find the condition to maintain the strong hyperbolicity in \cite{Beyer:2004sv}. The differences between  general relativity and $f(R)$ gravity are given by 
additional conditions $h\neq1$ and $GH\neq3/4$.  These conditons are peculiar in $f(R)$ gravity. 

The conditions to keep the strong hyperbolicity for the ADM formulation with modified harmonic gauge condition are $F\neq p$, $F\neq1$, $p\neq1$, $F>0$ and $p>0$.
In general relativity, it turns out the principal matrix $\mathbf{P}$ is diagonalizable if and only if $F>0$ and $F\neq1$~\cite{Sarbach:2012pr}. Therefore, among these conditions, $F\neq p$ and $p\neq1$ are more important for $f(R)$ gravity. Since we have $F\neq p$, in some sense, it means that the gauge equations for the lapse and shift function has to be scaled differently . 

In last formulation (which  can be called a generalized Z4 formulation),  a 4-vector $\mathcal{Z}^a$ has been added.  
Interestingly,  in $f(R)$ gravity, we find that the strong hyperbolicity can not be kept  if one writes the Z4 formulation as  the one in general relativity. 
Hence, the Z4 formulation here expressed as in Eq.(\ref{EOM_Z4}) with $l_1\neq l_2$ plays a vital role in the proof of the strong hyperbolicity. In a sort of sense, it is a correct Z4 formulation for the $f(R)$ gravity.

\section*{Acknowledgement}
We would like to thank Yaqi Zhao for her useful discussion during the early stage of this work. This work was supported in part by the National Natural Science Foundation of China with grants No.11622543, No.12075232, No.11947301, and No.12047502. This work is also supported by the Fundamental Research Funds for the
Central Universities under Grant No: WK2030000036, and the Key Research Program of the Chinese Academy of Sciences, Grant NO. XDPB15.

\bibliography{reference}{}

\begin{thebibliography}{45}%
\makeatletter
\providecommand \@ifxundefined [1]{%
 \@ifx{#1\undefined}
}%
\providecommand \@ifnum [1]{%
 \ifnum #1\expandafter \@firstoftwo
 \else \expandafter \@secondoftwo
 \fi
}%
\providecommand \@ifx [1]{%
 \ifx #1\expandafter \@firstoftwo
 \else \expandafter \@secondoftwo
 \fi
}%
\providecommand \natexlab [1]{#1}%
\providecommand \enquote  [1]{``#1''}%
\providecommand \bibnamefont  [1]{#1}%
\providecommand \bibfnamefont [1]{#1}%
\providecommand \citenamefont [1]{#1}%
\providecommand \href@noop [0]{\@secondoftwo}%
\providecommand \href [0]{\begingroup \@sanitize@url \@href}%
\providecommand \@href[1]{\@@startlink{#1}\@@href}%
\providecommand \@@href[1]{\endgroup#1\@@endlink}%
\providecommand \@sanitize@url [0]{\catcode `\\12\catcode `\$12\catcode
  `\&12\catcode `\#12\catcode `\^12\catcode `\_12\catcode `\%12\relax}%
\providecommand \@@startlink[1]{}%
\providecommand \@@endlink[0]{}%
\providecommand \url  [0]{\begingroup\@sanitize@url \@url }%
\providecommand \@url [1]{\endgroup\@href {#1}{\urlprefix }}%
\providecommand \urlprefix  [0]{URL }%
\providecommand \Eprint [0]{\href }%
\providecommand \doibase [0]{http://dx.doi.org/}%
\providecommand \selectlanguage [0]{\@gobble}%
\providecommand \bibinfo  [0]{\@secondoftwo}%
\providecommand \bibfield  [0]{\@secondoftwo}%
\providecommand \translation [1]{[#1]}%
\providecommand \BibitemOpen [0]{}%
\providecommand \bibitemStop [0]{}%
\providecommand \bibitemNoStop [0]{.\EOS\space}%
\providecommand \EOS [0]{\spacefactor3000\relax}%
\providecommand \BibitemShut  [1]{\csname bibitem#1\endcsname}%
\let\auto@bib@innerbib\@empty
\bibitem [{\citenamefont {Starobinsky}(1980)}]{Starobinsky:1980te}%
  \BibitemOpen
  \bibfield  {author} {\bibinfo {author} {\bibfnamefont {A.~A.}\ \bibnamefont
  {Starobinsky}},\ }\href {\doibase 10.1016/0370-2693(80)90670-X} {\bibfield
  {journal} {\bibinfo  {journal} {Phys. Lett. B}\ }\textbf {\bibinfo {volume}
  {91}},\ \bibinfo {pages} {99} (\bibinfo {year} {1980})}\BibitemShut {NoStop}%
\bibitem [{\citenamefont {Li}\ and\ \citenamefont {Barrow}(2007)}]{Li:2007xn}%
  \BibitemOpen
  \bibfield  {author} {\bibinfo {author} {\bibfnamefont {B.}~\bibnamefont
  {Li}}\ and\ \bibinfo {author} {\bibfnamefont {J.~D.}\ \bibnamefont
  {Barrow}},\ }\href {\doibase 10.1103/PhysRevD.75.084010} {\bibfield
  {journal} {\bibinfo  {journal} {Phys. Rev. D}\ }\textbf {\bibinfo {volume}
  {75}},\ \bibinfo {pages} {084010} (\bibinfo {year} {2007})},\ \Eprint
  {http://arxiv.org/abs/gr-qc/0701111} {arXiv:gr-qc/0701111} \BibitemShut
  {NoStop}%
\bibitem [{\citenamefont {Lecian}\ and\ \citenamefont
  {Montani}(2009)}]{Lecian:2008vc}%
  \BibitemOpen
  \bibfield  {author} {\bibinfo {author} {\bibfnamefont {O.~M.}\ \bibnamefont
  {Lecian}}\ and\ \bibinfo {author} {\bibfnamefont {G.}~\bibnamefont
  {Montani}},\ }\href {\doibase 10.1088/0264-9381/26/4/045014} {\bibfield
  {journal} {\bibinfo  {journal} {Class. Quant. Grav.}\ }\textbf {\bibinfo
  {volume} {26}},\ \bibinfo {pages} {045014} (\bibinfo {year} {2009})},\
  \Eprint {http://arxiv.org/abs/0807.4428} {arXiv:0807.4428 [gr-qc]}
  \BibitemShut {NoStop}%
\bibitem [{\citenamefont {Sotiriou}\ and\ \citenamefont
  {Faraoni}(2010)}]{Sotiriou:2008rp}%
  \BibitemOpen
  \bibfield  {author} {\bibinfo {author} {\bibfnamefont {T.~P.}\ \bibnamefont
  {Sotiriou}}\ and\ \bibinfo {author} {\bibfnamefont {V.}~\bibnamefont
  {Faraoni}},\ }\href {\doibase 10.1103/RevModPhys.82.451} {\bibfield
  {journal} {\bibinfo  {journal} {Rev. Mod. Phys.}\ }\textbf {\bibinfo {volume}
  {82}},\ \bibinfo {pages} {451} (\bibinfo {year} {2010})},\ \Eprint
  {http://arxiv.org/abs/0805.1726} {arXiv:0805.1726 [gr-qc]} \BibitemShut
  {NoStop}%
\bibitem [{\citenamefont {De~Felice}\ and\ \citenamefont
  {Tsujikawa}(2010)}]{DeFelice:2010aj}%
  \BibitemOpen
  \bibfield  {author} {\bibinfo {author} {\bibfnamefont {A.}~\bibnamefont
  {De~Felice}}\ and\ \bibinfo {author} {\bibfnamefont {S.}~\bibnamefont
  {Tsujikawa}},\ }\href {\doibase 10.12942/lrr-2010-3} {\bibfield  {journal}
  {\bibinfo  {journal} {Living Rev. Rel.}\ }\textbf {\bibinfo {volume} {13}},\
  \bibinfo {pages} {3} (\bibinfo {year} {2010})},\ \Eprint
  {http://arxiv.org/abs/1002.4928} {arXiv:1002.4928 [gr-qc]} \BibitemShut
  {NoStop}%
\bibitem [{\citenamefont {Salgado}\ \emph {et~al.}(2008)\citenamefont
  {Salgado}, \citenamefont {Martinez-del Rio}, \citenamefont {Alcubierre},\
  and\ \citenamefont {Nunez}}]{Salgado:2008xh}%
  \BibitemOpen
  \bibfield  {author} {\bibinfo {author} {\bibfnamefont {M.}~\bibnamefont
  {Salgado}}, \bibinfo {author} {\bibfnamefont {D.}~\bibnamefont {Martinez-del
  Rio}}, \bibinfo {author} {\bibfnamefont {M.}~\bibnamefont {Alcubierre}}, \
  and\ \bibinfo {author} {\bibfnamefont {D.}~\bibnamefont {Nunez}},\ }\href
  {\doibase 10.1103/PhysRevD.77.104010} {\bibfield  {journal} {\bibinfo
  {journal} {Phys. Rev. D}\ }\textbf {\bibinfo {volume} {77}},\ \bibinfo
  {pages} {104010} (\bibinfo {year} {2008})},\ \Eprint
  {http://arxiv.org/abs/0801.2372} {arXiv:0801.2372 [gr-qc]} \BibitemShut
  {NoStop}%
\bibitem [{\citenamefont {Sarbach}\ \emph {et~al.}(2019)\citenamefont
  {Sarbach}, \citenamefont {Barausse},\ and\ \citenamefont
  {Preciado-L\'opez}}]{Sarbach:2019yso}%
  \BibitemOpen
  \bibfield  {author} {\bibinfo {author} {\bibfnamefont {O.}~\bibnamefont
  {Sarbach}}, \bibinfo {author} {\bibfnamefont {E.}~\bibnamefont {Barausse}}, \
  and\ \bibinfo {author} {\bibfnamefont {J.~A.}\ \bibnamefont
  {Preciado-L\'opez}},\ }\href {\doibase 10.1088/1361-6382/ab2e13} {\bibfield
  {journal} {\bibinfo  {journal} {Class. Quant. Grav.}\ }\textbf {\bibinfo
  {volume} {36}},\ \bibinfo {pages} {165007} (\bibinfo {year} {2019})},\
  \Eprint {http://arxiv.org/abs/1902.05130} {arXiv:1902.05130 [gr-qc]}
  \BibitemShut {NoStop}%
\bibitem [{\citenamefont {Buchman}\ and\ \citenamefont
  {Bardeen}(2003)}]{Buchman:2003sq}%
  \BibitemOpen
  \bibfield  {author} {\bibinfo {author} {\bibfnamefont {L.~T.}\ \bibnamefont
  {Buchman}}\ and\ \bibinfo {author} {\bibfnamefont {J.~M.}\ \bibnamefont
  {Bardeen}},\ }\href {\doibase 10.1103/PhysRevD.72.049903} {\bibfield
  {journal} {\bibinfo  {journal} {Phys. Rev. D}\ }\textbf {\bibinfo {volume}
  {67}},\ \bibinfo {pages} {084017} (\bibinfo {year} {2003})},\ \bibinfo {note}
  {[Erratum: Phys.Rev.D 72, 049903 (2005)]},\ \Eprint
  {http://arxiv.org/abs/gr-qc/0301072} {arXiv:gr-qc/0301072} \BibitemShut
  {NoStop}%
\bibitem [{\citenamefont {Kov\'acs}(2019)}]{Kovacs:2019jqj}%
  \BibitemOpen
  \bibfield  {author} {\bibinfo {author} {\bibfnamefont {A.~D.}\ \bibnamefont
  {Kov\'acs}},\ }\href {\doibase 10.1103/PhysRevD.100.024005} {\bibfield
  {journal} {\bibinfo  {journal} {Phys. Rev. D}\ }\textbf {\bibinfo {volume}
  {100}},\ \bibinfo {pages} {024005} (\bibinfo {year} {2019})},\ \Eprint
  {http://arxiv.org/abs/1904.00963} {arXiv:1904.00963 [gr-qc]} \BibitemShut
  {NoStop}%
\bibitem [{\citenamefont {Taylor}(1981)}]{Taylor1}%
  \BibitemOpen
  \bibfield  {author} {\bibinfo {author} {\bibfnamefont {M.~E.}\ \bibnamefont
  {Taylor}},\ }\href@noop {} {\emph {\bibinfo {title} {{Pseudodifferential
  operators}}}}\ (\bibinfo  {publisher} {Princeton University Press, Princeton,
  New Jersey},\ \bibinfo {year} {1981})\BibitemShut {NoStop}%
\bibitem [{\citenamefont {Taylor}(1991)}]{Taylor2}%
  \BibitemOpen
  \bibfield  {author} {\bibinfo {author} {\bibfnamefont {M.~E.}\ \bibnamefont
  {Taylor}},\ }\href@noop {} {\emph {\bibinfo {title} {{Progress in
  Mathematics}}}},\ Vol.\ \bibinfo {volume} {100}\ (\bibinfo  {publisher}
  {Birkhauser, Boston-Basel-Berlin},\ \bibinfo {year} {1991})\BibitemShut
  {NoStop}%
\bibitem [{\citenamefont {Papallo}\ and\ \citenamefont
  {Reall}(2017)}]{Papallo:2017qvl}%
  \BibitemOpen
  \bibfield  {author} {\bibinfo {author} {\bibfnamefont {G.}~\bibnamefont
  {Papallo}}\ and\ \bibinfo {author} {\bibfnamefont {H.~S.}\ \bibnamefont
  {Reall}},\ }\href {\doibase 10.1103/PhysRevD.96.044019} {\bibfield  {journal}
  {\bibinfo  {journal} {Phys. Rev. D}\ }\textbf {\bibinfo {volume} {96}},\
  \bibinfo {pages} {044019} (\bibinfo {year} {2017})},\ \Eprint
  {http://arxiv.org/abs/1705.04370} {arXiv:1705.04370 [gr-qc]} \BibitemShut
  {NoStop}%
\bibitem [{\citenamefont {Papallo}(2017)}]{Papallo:2017ddx}%
  \BibitemOpen
  \bibfield  {author} {\bibinfo {author} {\bibfnamefont {G.}~\bibnamefont
  {Papallo}},\ }\href {\doibase 10.1103/PhysRevD.96.124036} {\bibfield
  {journal} {\bibinfo  {journal} {Phys. Rev. D}\ }\textbf {\bibinfo {volume}
  {96}},\ \bibinfo {pages} {124036} (\bibinfo {year} {2017})},\ \Eprint
  {http://arxiv.org/abs/1710.10155} {arXiv:1710.10155 [gr-qc]} \BibitemShut
  {NoStop}%
\bibitem [{\citenamefont {Kov\'acs}\ and\ \citenamefont
  {Reall}(2020)}]{Kovacs:2020ywu}%
  \BibitemOpen
  \bibfield  {author} {\bibinfo {author} {\bibfnamefont {A.~D.}\ \bibnamefont
  {Kov\'acs}}\ and\ \bibinfo {author} {\bibfnamefont {H.~S.}\ \bibnamefont
  {Reall}},\ }\href {\doibase 10.1103/PhysRevD.101.124003} {\bibfield
  {journal} {\bibinfo  {journal} {Phys. Rev. D}\ }\textbf {\bibinfo {volume}
  {101}},\ \bibinfo {pages} {124003} (\bibinfo {year} {2020})},\ \Eprint
  {http://arxiv.org/abs/2003.08398} {arXiv:2003.08398 [gr-qc]} \BibitemShut
  {NoStop}%
\bibitem [{\citenamefont {Cao}\ and\ \citenamefont
  {Wu}(2021{\natexlab{a}})}]{Cao:2021sty}%
  \BibitemOpen
  \bibfield  {author} {\bibinfo {author} {\bibfnamefont {L.-M.}\ \bibnamefont
  {Cao}}\ and\ \bibinfo {author} {\bibfnamefont {L.-B.}\ \bibnamefont {Wu}},\
  }\href {\doibase 10.1103/PhysRevD.103.064054} {\bibfield  {journal} {\bibinfo
   {journal} {Phys. Rev. D}\ }\textbf {\bibinfo {volume} {103}},\ \bibinfo
  {pages} {064054} (\bibinfo {year} {2021}{\natexlab{a}})},\ \Eprint
  {http://arxiv.org/abs/2101.02461} {arXiv:2101.02461 [gr-qc]} \BibitemShut
  {NoStop}%
\bibitem [{\citenamefont {Cao}\ and\ \citenamefont
  {Wu}(2021{\natexlab{b}})}]{Cao:2021nng}%
  \BibitemOpen
  \bibfield  {author} {\bibinfo {author} {\bibfnamefont {L.-M.}\ \bibnamefont
  {Cao}}\ and\ \bibinfo {author} {\bibfnamefont {L.-B.}\ \bibnamefont {Wu}},\
  }\href@noop {} {\  (\bibinfo {year} {2021}{\natexlab{b}})},\ \Eprint
  {http://arxiv.org/abs/2103.09612} {arXiv:2103.09612 [gr-qc]} \BibitemShut
  {NoStop}%
\bibitem [{\citenamefont {Arnowitt}\ \emph {et~al.}(2008)\citenamefont
  {Arnowitt}, \citenamefont {Deser},\ and\ \citenamefont
  {Misner}}]{Arnowitt:1962hi}%
  \BibitemOpen
  \bibfield  {author} {\bibinfo {author} {\bibfnamefont {R.~L.}\ \bibnamefont
  {Arnowitt}}, \bibinfo {author} {\bibfnamefont {S.}~\bibnamefont {Deser}}, \
  and\ \bibinfo {author} {\bibfnamefont {C.~W.}\ \bibnamefont {Misner}},\
  }\href {\doibase 10.1007/s10714-008-0661-1} {\bibfield  {journal} {\bibinfo
  {journal} {Gen. Rel. Grav.}\ }\textbf {\bibinfo {volume} {40}},\ \bibinfo
  {pages} {1997} (\bibinfo {year} {2008})},\ \Eprint
  {http://arxiv.org/abs/gr-qc/0405109} {arXiv:gr-qc/0405109} \BibitemShut
  {NoStop}%
\bibitem [{\citenamefont {Bondi}\ \emph {et~al.}(1962)\citenamefont {Bondi},
  \citenamefont {van~der Burg},\ and\ \citenamefont {Metzner}}]{Bondi:1962px}%
  \BibitemOpen
  \bibfield  {author} {\bibinfo {author} {\bibfnamefont {H.}~\bibnamefont
  {Bondi}}, \bibinfo {author} {\bibfnamefont {M.~G.~J.}\ \bibnamefont {van~der
  Burg}}, \ and\ \bibinfo {author} {\bibfnamefont {A.~W.~K.}\ \bibnamefont
  {Metzner}},\ }\href {\doibase 10.1098/rspa.1962.0161} {\bibfield  {journal}
  {\bibinfo  {journal} {Proc. Roy. Soc. Lond. A}\ }\textbf {\bibinfo {volume}
  {269}},\ \bibinfo {pages} {21} (\bibinfo {year} {1962})}\BibitemShut
  {NoStop}%
\bibitem [{\citenamefont {Winicour}(2009)}]{Winicour:2008vpn}%
  \BibitemOpen
  \bibfield  {author} {\bibinfo {author} {\bibfnamefont {J.}~\bibnamefont
  {Winicour}},\ }\href {\doibase 10.12942/lrr-2009-3} {\bibfield  {journal}
  {\bibinfo  {journal} {Living Rev. Rel.}\ }\textbf {\bibinfo {volume} {12}},\
  \bibinfo {pages} {3} (\bibinfo {year} {2009})},\ \Eprint
  {http://arxiv.org/abs/0810.1903} {arXiv:0810.1903 [gr-qc]} \BibitemShut
  {NoStop}%
\bibitem [{\citenamefont {Giannakopoulos}\ \emph {et~al.}(2020)\citenamefont
  {Giannakopoulos}, \citenamefont {Hilditch},\ and\ \citenamefont
  {Zilhao}}]{Giannakopoulos:2020dih}%
  \BibitemOpen
  \bibfield  {author} {\bibinfo {author} {\bibfnamefont {T.}~\bibnamefont
  {Giannakopoulos}}, \bibinfo {author} {\bibfnamefont {D.}~\bibnamefont
  {Hilditch}}, \ and\ \bibinfo {author} {\bibfnamefont {M.}~\bibnamefont
  {Zilhao}},\ }\href {\doibase 10.1103/PhysRevD.102.064035} {\bibfield
  {journal} {\bibinfo  {journal} {Phys. Rev. D}\ }\textbf {\bibinfo {volume}
  {102}},\ \bibinfo {pages} {064035} (\bibinfo {year} {2020})},\ \Eprint
  {http://arxiv.org/abs/2007.06419} {arXiv:2007.06419 [gr-qc]} \BibitemShut
  {NoStop}%
\bibitem [{\citenamefont {Nagy}\ \emph {et~al.}(2004)\citenamefont {Nagy},
  \citenamefont {Ortiz},\ and\ \citenamefont {Reula}}]{Nagy:2004td}%
  \BibitemOpen
  \bibfield  {author} {\bibinfo {author} {\bibfnamefont {G.}~\bibnamefont
  {Nagy}}, \bibinfo {author} {\bibfnamefont {O.~E.}\ \bibnamefont {Ortiz}}, \
  and\ \bibinfo {author} {\bibfnamefont {O.~A.}\ \bibnamefont {Reula}},\ }\href
  {\doibase 10.1103/PhysRevD.70.044012} {\bibfield  {journal} {\bibinfo
  {journal} {Phys. Rev. D}\ }\textbf {\bibinfo {volume} {70}},\ \bibinfo
  {pages} {044012} (\bibinfo {year} {2004})},\ \Eprint
  {http://arxiv.org/abs/gr-qc/0402123} {arXiv:gr-qc/0402123} \BibitemShut
  {NoStop}%
\bibitem [{\citenamefont {Sarbach}\ and\ \citenamefont
  {Tiglio}(2012)}]{Sarbach:2012pr}%
  \BibitemOpen
  \bibfield  {author} {\bibinfo {author} {\bibfnamefont {O.}~\bibnamefont
  {Sarbach}}\ and\ \bibinfo {author} {\bibfnamefont {M.}~\bibnamefont
  {Tiglio}},\ }\href {\doibase 10.12942/lrr-2012-9} {\bibfield  {journal}
  {\bibinfo  {journal} {Living Rev. Rel.}\ }\textbf {\bibinfo {volume} {15}},\
  \bibinfo {pages} {9} (\bibinfo {year} {2012})},\ \Eprint
  {http://arxiv.org/abs/1203.6443} {arXiv:1203.6443 [gr-qc]} \BibitemShut
  {NoStop}%
\bibitem [{\citenamefont {Kreiss}\ and\ \citenamefont
  {Lorenz}(1989)}]{1989Initial}%
  \BibitemOpen
  \bibfield  {author} {\bibinfo {author} {\bibfnamefont {H.~O.}\ \bibnamefont
  {Kreiss}}\ and\ \bibinfo {author} {\bibfnamefont {J.}~\bibnamefont
  {Lorenz}},\ }\href@noop {} {\bibfield  {journal} {\bibinfo  {journal}
  {Academic Press}\ } (\bibinfo {year} {1989})}\BibitemShut {NoStop}%
\bibitem [{\citenamefont {Kidder}\ \emph {et~al.}(2001)\citenamefont {Kidder},
  \citenamefont {Scheel},\ and\ \citenamefont {Teukolsky}}]{Kidder:2001tz}%
  \BibitemOpen
  \bibfield  {author} {\bibinfo {author} {\bibfnamefont {L.~E.}\ \bibnamefont
  {Kidder}}, \bibinfo {author} {\bibfnamefont {M.~A.}\ \bibnamefont {Scheel}},
  \ and\ \bibinfo {author} {\bibfnamefont {S.~A.}\ \bibnamefont {Teukolsky}},\
  }\href {\doibase 10.1103/PhysRevD.64.064017} {\bibfield  {journal} {\bibinfo
  {journal} {Phys. Rev. D}\ }\textbf {\bibinfo {volume} {64}},\ \bibinfo
  {pages} {064017} (\bibinfo {year} {2001})},\ \Eprint
  {http://arxiv.org/abs/gr-qc/0105031} {arXiv:gr-qc/0105031} \BibitemShut
  {NoStop}%
\bibitem [{\citenamefont {Calabrese}\ \emph
  {et~al.}(2002{\natexlab{a}})\citenamefont {Calabrese}, \citenamefont
  {Pullin}, \citenamefont {Sarbach},\ and\ \citenamefont
  {Tiglio}}]{Calabrese:2002ej}%
  \BibitemOpen
  \bibfield  {author} {\bibinfo {author} {\bibfnamefont {G.}~\bibnamefont
  {Calabrese}}, \bibinfo {author} {\bibfnamefont {J.}~\bibnamefont {Pullin}},
  \bibinfo {author} {\bibfnamefont {O.}~\bibnamefont {Sarbach}}, \ and\
  \bibinfo {author} {\bibfnamefont {M.}~\bibnamefont {Tiglio}},\ }\href
  {\doibase 10.1103/PhysRevD.66.041501} {\bibfield  {journal} {\bibinfo
  {journal} {Phys. Rev. D}\ }\textbf {\bibinfo {volume} {66}},\ \bibinfo
  {pages} {041501} (\bibinfo {year} {2002}{\natexlab{a}})},\ \Eprint
  {http://arxiv.org/abs/gr-qc/0207018} {arXiv:gr-qc/0207018} \BibitemShut
  {NoStop}%
\bibitem [{\citenamefont {Calabrese}\ \emph
  {et~al.}(2002{\natexlab{b}})\citenamefont {Calabrese}, \citenamefont
  {Pullin}, \citenamefont {Sarbach},\ and\ \citenamefont
  {Tiglio}}]{Calabrese:2002ei}%
  \BibitemOpen
  \bibfield  {author} {\bibinfo {author} {\bibfnamefont {G.}~\bibnamefont
  {Calabrese}}, \bibinfo {author} {\bibfnamefont {J.}~\bibnamefont {Pullin}},
  \bibinfo {author} {\bibfnamefont {O.}~\bibnamefont {Sarbach}}, \ and\
  \bibinfo {author} {\bibfnamefont {M.}~\bibnamefont {Tiglio}},\ }\href
  {\doibase 10.1103/PhysRevD.66.064011} {\bibfield  {journal} {\bibinfo
  {journal} {Phys. Rev. D}\ }\textbf {\bibinfo {volume} {66}},\ \bibinfo
  {pages} {064011} (\bibinfo {year} {2002}{\natexlab{b}})},\ \Eprint
  {http://arxiv.org/abs/gr-qc/0205073} {arXiv:gr-qc/0205073} \BibitemShut
  {NoStop}%
\bibitem [{\citenamefont {Baumgarte}\ and\ \citenamefont
  {Shapiro}(1998)}]{Baumgarte:1998te}%
  \BibitemOpen
  \bibfield  {author} {\bibinfo {author} {\bibfnamefont {T.~W.}\ \bibnamefont
  {Baumgarte}}\ and\ \bibinfo {author} {\bibfnamefont {S.~L.}\ \bibnamefont
  {Shapiro}},\ }\href {\doibase 10.1103/PhysRevD.59.024007} {\bibfield
  {journal} {\bibinfo  {journal} {Phys. Rev. D}\ }\textbf {\bibinfo {volume}
  {59}},\ \bibinfo {pages} {024007} (\bibinfo {year} {1998})},\ \Eprint
  {http://arxiv.org/abs/gr-qc/9810065} {arXiv:gr-qc/9810065} \BibitemShut
  {NoStop}%
\bibitem [{\citenamefont {Sarbach}\ \emph {et~al.}(2002)\citenamefont
  {Sarbach}, \citenamefont {Calabrese}, \citenamefont {Pullin},\ and\
  \citenamefont {Tiglio}}]{Sarbach:2002bt}%
  \BibitemOpen
  \bibfield  {author} {\bibinfo {author} {\bibfnamefont {O.}~\bibnamefont
  {Sarbach}}, \bibinfo {author} {\bibfnamefont {G.}~\bibnamefont {Calabrese}},
  \bibinfo {author} {\bibfnamefont {J.}~\bibnamefont {Pullin}}, \ and\ \bibinfo
  {author} {\bibfnamefont {M.}~\bibnamefont {Tiglio}},\ }\href {\doibase
  10.1103/PhysRevD.66.064002} {\bibfield  {journal} {\bibinfo  {journal} {Phys.
  Rev. D}\ }\textbf {\bibinfo {volume} {66}},\ \bibinfo {pages} {064002}
  (\bibinfo {year} {2002})},\ \Eprint {http://arxiv.org/abs/gr-qc/0205064}
  {arXiv:gr-qc/0205064} \BibitemShut {NoStop}%
\bibitem [{\citenamefont {Beyer}\ and\ \citenamefont
  {Sarbach}(2004)}]{Beyer:2004sv}%
  \BibitemOpen
  \bibfield  {author} {\bibinfo {author} {\bibfnamefont {H.~R.}\ \bibnamefont
  {Beyer}}\ and\ \bibinfo {author} {\bibfnamefont {O.}~\bibnamefont
  {Sarbach}},\ }\href {\doibase 10.1103/PhysRevD.70.104004} {\bibfield
  {journal} {\bibinfo  {journal} {Phys. Rev. D}\ }\textbf {\bibinfo {volume}
  {70}},\ \bibinfo {pages} {104004} (\bibinfo {year} {2004})},\ \Eprint
  {http://arxiv.org/abs/gr-qc/0406003} {arXiv:gr-qc/0406003} \BibitemShut
  {NoStop}%
\bibitem [{\citenamefont {Reula}(2004)}]{Reula:2004xd}%
  \BibitemOpen
  \bibfield  {author} {\bibinfo {author} {\bibfnamefont {O.~A.}\ \bibnamefont
  {Reula}},\ }\href {\doibase 10.1142/S0219891604000111} {\bibfield  {journal}
  {\bibinfo  {journal} {J. Hyperbol. Diff. Equat.}\ }\textbf {\bibinfo {volume}
  {1}},\ \bibinfo {pages} {251} (\bibinfo {year} {2004})},\ \Eprint
  {http://arxiv.org/abs/gr-qc/0403007} {arXiv:gr-qc/0403007} \BibitemShut
  {NoStop}%
\bibitem [{\citenamefont {Frittelli}\ and\ \citenamefont
  {Reula}(1999)}]{Frittelli:1999sj}%
  \BibitemOpen
  \bibfield  {author} {\bibinfo {author} {\bibfnamefont {S.}~\bibnamefont
  {Frittelli}}\ and\ \bibinfo {author} {\bibfnamefont {O.~A.}\ \bibnamefont
  {Reula}},\ }\href {\doibase 10.1063/1.533022} {\bibfield  {journal} {\bibinfo
   {journal} {J. Math. Phys.}\ }\textbf {\bibinfo {volume} {40}},\ \bibinfo
  {pages} {5143} (\bibinfo {year} {1999})},\ \Eprint
  {http://arxiv.org/abs/gr-qc/9904048} {arXiv:gr-qc/9904048} \BibitemShut
  {NoStop}%
\bibitem [{\citenamefont {Kreiss}\ and\ \citenamefont {Ortiz}(2002)}]{Kriess}%
  \BibitemOpen
  \bibfield  {author} {\bibinfo {author} {\bibfnamefont {H.}~\bibnamefont
  {Kreiss}}\ and\ \bibinfo {author} {\bibfnamefont {O.~E.}\ \bibnamefont
  {Ortiz}},\ }\href@noop {} {\emph {\bibinfo {title} {The conformal structure
  of spacetimes: Geometry, Analysis, Numerics.}}},\ Vol.\ \bibinfo {volume}
  {604}\ (\bibinfo  {publisher} {Springer Lecture Notes in Physics},\ \bibinfo
  {address} {Heidelberg},\ \bibinfo {year} {2002})\BibitemShut {NoStop}%
\bibitem [{\citenamefont {Lanahan-Tremblay}\ and\ \citenamefont
  {Faraoni}(2007)}]{Lanahan-Tremblay:2007sxd}%
  \BibitemOpen
  \bibfield  {author} {\bibinfo {author} {\bibfnamefont {N.}~\bibnamefont
  {Lanahan-Tremblay}}\ and\ \bibinfo {author} {\bibfnamefont {V.}~\bibnamefont
  {Faraoni}},\ }\href {\doibase 10.1088/0264-9381/24/22/024} {\bibfield
  {journal} {\bibinfo  {journal} {Class. Quant. Grav.}\ }\textbf {\bibinfo
  {volume} {24}},\ \bibinfo {pages} {5667} (\bibinfo {year} {2007})},\ \Eprint
  {http://arxiv.org/abs/0709.4414} {arXiv:0709.4414 [gr-qc]} \BibitemShut
  {NoStop}%
\bibitem [{\citenamefont {Tsokaros}(2014)}]{Tsokaros:2013fma}%
  \BibitemOpen
  \bibfield  {author} {\bibinfo {author} {\bibfnamefont {A.}~\bibnamefont
  {Tsokaros}},\ }\href {\doibase 10.1088/0264-9381/31/2/025021} {\bibfield
  {journal} {\bibinfo  {journal} {Class. Quant. Grav.}\ }\textbf {\bibinfo
  {volume} {31}},\ \bibinfo {pages} {025021} (\bibinfo {year}
  {2014})}\BibitemShut {NoStop}%
\bibitem [{\citenamefont {Mongwane}(2016)}]{Mongwane:2016qtz}%
  \BibitemOpen
  \bibfield  {author} {\bibinfo {author} {\bibfnamefont {B.}~\bibnamefont
  {Mongwane}},\ }\href {\doibase 10.1007/s10714-016-2147-x} {\bibfield
  {journal} {\bibinfo  {journal} {Gen. Rel. Grav.}\ }\textbf {\bibinfo {volume}
  {48}},\ \bibinfo {pages} {152} (\bibinfo {year} {2016})},\ \Eprint
  {http://arxiv.org/abs/1610.07224} {arXiv:1610.07224 [gr-qc]} \BibitemShut
  {NoStop}%
\bibitem [{\citenamefont {Bona}\ \emph {et~al.}(1995)\citenamefont {Bona},
  \citenamefont {Masso}, \citenamefont {Seidel},\ and\ \citenamefont
  {Stela}}]{Bona:1994dr}%
  \BibitemOpen
  \bibfield  {author} {\bibinfo {author} {\bibfnamefont {C.}~\bibnamefont
  {Bona}}, \bibinfo {author} {\bibfnamefont {J.}~\bibnamefont {Masso}},
  \bibinfo {author} {\bibfnamefont {E.}~\bibnamefont {Seidel}}, \ and\ \bibinfo
  {author} {\bibfnamefont {J.}~\bibnamefont {Stela}},\ }\href {\doibase
  10.1103/PhysRevLett.75.600} {\bibfield  {journal} {\bibinfo  {journal} {Phys.
  Rev. Lett.}\ }\textbf {\bibinfo {volume} {75}},\ \bibinfo {pages} {600}
  (\bibinfo {year} {1995})},\ \Eprint {http://arxiv.org/abs/gr-qc/9412071}
  {arXiv:gr-qc/9412071} \BibitemShut {NoStop}%
\bibitem [{\citenamefont {Sotiriou}(2006)}]{Sotiriou:2006hs}%
  \BibitemOpen
  \bibfield  {author} {\bibinfo {author} {\bibfnamefont {T.~P.}\ \bibnamefont
  {Sotiriou}},\ }\href {\doibase 10.1088/0264-9381/23/17/003} {\bibfield
  {journal} {\bibinfo  {journal} {Class. Quant. Grav.}\ }\textbf {\bibinfo
  {volume} {23}},\ \bibinfo {pages} {5117} (\bibinfo {year} {2006})},\ \Eprint
  {http://arxiv.org/abs/gr-qc/0604028} {arXiv:gr-qc/0604028} \BibitemShut
  {NoStop}%
\bibitem [{\citenamefont {Alcubierre}\ \emph {et~al.}(2000)\citenamefont
  {Alcubierre}, \citenamefont {Allen}, \citenamefont {Bruegmann}, \citenamefont
  {Seidel},\ and\ \citenamefont {Suen}}]{Alcubierre:1999rt}%
  \BibitemOpen
  \bibfield  {author} {\bibinfo {author} {\bibfnamefont {M.}~\bibnamefont
  {Alcubierre}}, \bibinfo {author} {\bibfnamefont {G.}~\bibnamefont {Allen}},
  \bibinfo {author} {\bibfnamefont {B.}~\bibnamefont {Bruegmann}}, \bibinfo
  {author} {\bibfnamefont {E.}~\bibnamefont {Seidel}}, \ and\ \bibinfo {author}
  {\bibfnamefont {W.-M.}\ \bibnamefont {Suen}},\ }\href {\doibase
  10.1103/PhysRevD.62.124011} {\bibfield  {journal} {\bibinfo  {journal} {Phys.
  Rev. D}\ }\textbf {\bibinfo {volume} {62}},\ \bibinfo {pages} {124011}
  (\bibinfo {year} {2000})},\ \Eprint {http://arxiv.org/abs/gr-qc/9908079}
  {arXiv:gr-qc/9908079} \BibitemShut {NoStop}%
\bibitem [{\citenamefont {Alcubierre}\ \emph {et~al.}(2003)\citenamefont
  {Alcubierre}, \citenamefont {Bruegmann}, \citenamefont {Diener},
  \citenamefont {Koppitz}, \citenamefont {Pollney}, \citenamefont {Seidel},\
  and\ \citenamefont {Takahashi}}]{Alcubierre:2002kk}%
  \BibitemOpen
  \bibfield  {author} {\bibinfo {author} {\bibfnamefont {M.}~\bibnamefont
  {Alcubierre}}, \bibinfo {author} {\bibfnamefont {B.}~\bibnamefont
  {Bruegmann}}, \bibinfo {author} {\bibfnamefont {P.}~\bibnamefont {Diener}},
  \bibinfo {author} {\bibfnamefont {M.}~\bibnamefont {Koppitz}}, \bibinfo
  {author} {\bibfnamefont {D.}~\bibnamefont {Pollney}}, \bibinfo {author}
  {\bibfnamefont {E.}~\bibnamefont {Seidel}}, \ and\ \bibinfo {author}
  {\bibfnamefont {R.}~\bibnamefont {Takahashi}},\ }\href {\doibase
  10.1103/PhysRevD.67.084023} {\bibfield  {journal} {\bibinfo  {journal} {Phys.
  Rev. D}\ }\textbf {\bibinfo {volume} {67}},\ \bibinfo {pages} {084023}
  (\bibinfo {year} {2003})},\ \Eprint {http://arxiv.org/abs/gr-qc/0206072}
  {arXiv:gr-qc/0206072} \BibitemShut {NoStop}%
\bibitem [{\citenamefont {Friedrich}(1996)}]{Friedrich:1996hq}%
  \BibitemOpen
  \bibfield  {author} {\bibinfo {author} {\bibfnamefont {H.}~\bibnamefont
  {Friedrich}},\ }\href {\doibase 10.1088/0264-9381/13/6/014} {\bibfield
  {journal} {\bibinfo  {journal} {Class. Quant. Grav.}\ }\textbf {\bibinfo
  {volume} {13}},\ \bibinfo {pages} {1451} (\bibinfo {year}
  {1996})}\BibitemShut {NoStop}%
\bibitem [{\citenamefont {Garfinkle}(2002)}]{Garfinkle:2001ni}%
  \BibitemOpen
  \bibfield  {author} {\bibinfo {author} {\bibfnamefont {D.}~\bibnamefont
  {Garfinkle}},\ }\href {\doibase 10.1103/PhysRevD.65.044029} {\bibfield
  {journal} {\bibinfo  {journal} {Phys. Rev. D}\ }\textbf {\bibinfo {volume}
  {65}},\ \bibinfo {pages} {044029} (\bibinfo {year} {2002})},\ \Eprint
  {http://arxiv.org/abs/gr-qc/0110013} {arXiv:gr-qc/0110013} \BibitemShut
  {NoStop}%
\bibitem [{\citenamefont {Hawking}\ and\ \citenamefont
  {Ellis}(2011)}]{Hawking:1973uf}%
  \BibitemOpen
  \bibfield  {author} {\bibinfo {author} {\bibfnamefont {S.~W.}\ \bibnamefont
  {Hawking}}\ and\ \bibinfo {author} {\bibfnamefont {G.~F.~R.}\ \bibnamefont
  {Ellis}},\ }\href {\doibase 10.1017/CBO9780511524646} {\emph {\bibinfo
  {title} {{The Large Scale Structure of Space-Time}}}},\ Cambridge Monographs
  on Mathematical Physics\ (\bibinfo  {publisher} {Cambridge University
  Press},\ \bibinfo {year} {2011})\BibitemShut {NoStop}%
\bibitem [{\citenamefont {Bona}\ \emph {et~al.}(2002)\citenamefont {Bona},
  \citenamefont {Ledvinka},\ and\ \citenamefont {Palenzuela}}]{Bona:2002fq}%
  \BibitemOpen
  \bibfield  {author} {\bibinfo {author} {\bibfnamefont {C.}~\bibnamefont
  {Bona}}, \bibinfo {author} {\bibfnamefont {T.}~\bibnamefont {Ledvinka}}, \
  and\ \bibinfo {author} {\bibfnamefont {C.}~\bibnamefont {Palenzuela}},\
  }\href {\doibase 10.1103/PhysRevD.66.084013} {\bibfield  {journal} {\bibinfo
  {journal} {Phys. Rev. D}\ }\textbf {\bibinfo {volume} {66}},\ \bibinfo
  {pages} {084013} (\bibinfo {year} {2002})},\ \Eprint
  {http://arxiv.org/abs/gr-qc/0208087} {arXiv:gr-qc/0208087} \BibitemShut
  {NoStop}%
\bibitem [{\citenamefont {Bona}\ and\ \citenamefont
  {Palenzuela}(2004)}]{Bona:2004yp}%
  \BibitemOpen
  \bibfield  {author} {\bibinfo {author} {\bibfnamefont {C.}~\bibnamefont
  {Bona}}\ and\ \bibinfo {author} {\bibfnamefont {C.}~\bibnamefont
  {Palenzuela}},\ }\href {\doibase 10.1103/PhysRevD.69.104003} {\bibfield
  {journal} {\bibinfo  {journal} {Phys. Rev. D}\ }\textbf {\bibinfo {volume}
  {69}},\ \bibinfo {pages} {104003} (\bibinfo {year} {2004})},\ \Eprint
  {http://arxiv.org/abs/gr-qc/0401019} {arXiv:gr-qc/0401019} \BibitemShut
  {NoStop}%
\bibitem [{\citenamefont {Gundlach}\ \emph {et~al.}(2005)\citenamefont
  {Gundlach}, \citenamefont {Martin-Garcia}, \citenamefont {Calabrese},\ and\
  \citenamefont {Hinder}}]{Gundlach:2005eh}%
  \BibitemOpen
  \bibfield  {author} {\bibinfo {author} {\bibfnamefont {C.}~\bibnamefont
  {Gundlach}}, \bibinfo {author} {\bibfnamefont {J.~M.}\ \bibnamefont
  {Martin-Garcia}}, \bibinfo {author} {\bibfnamefont {G.}~\bibnamefont
  {Calabrese}}, \ and\ \bibinfo {author} {\bibfnamefont {I.}~\bibnamefont
  {Hinder}},\ }\href {\doibase 10.1088/0264-9381/22/17/025} {\bibfield
  {journal} {\bibinfo  {journal} {Class. Quant. Grav.}\ }\textbf {\bibinfo
  {volume} {22}},\ \bibinfo {pages} {3767} (\bibinfo {year} {2005})},\ \Eprint
  {http://arxiv.org/abs/gr-qc/0504114} {arXiv:gr-qc/0504114} \BibitemShut
  {NoStop}%
\end{thebibliography}%
\bibliographystyle{apsrev4-1}


\end{document}